\documentclass[traditabstract]{aa}

\usepackage{graphicx}
\usepackage[varg]{txfonts}
\usepackage{natbib}
\bibpunct{(}{)}{;}{a}{}{,}
\usepackage{array}
\usepackage{xspace}
\usepackage{lscape}
\usepackage{url}
\usepackage{arydshln}

\usepackage[hyperfootnotes=false, linktocpage=true, breaklinks=true, colorlinks=true, linkcolor=blue, citecolor=blue, urlcolor=blue]{hyperref}
\usepackage[all]{hypcap}

\newcommand{\FeKa}{Fe K\ensuremath{\alpha}\xspace}

\newcommand{\kms}{\ensuremath{\mathrm{km\ s^{-1}}}\xspace}
\newcommand{\NH}{\ensuremath{N_{\mathrm{H}}}\xspace}

\newcommand{\xmm}{{\it XMM-Newton}\xspace}
\newcommand{\chandra}{{\it Chandra}\xspace}
\newcommand{\swift}{{\it Swift}\xspace}
\newcommand{\exosat}{{EXOSAT}\xspace}

\newcommand{\integral}{{INTEGRAL}\xspace}
\newcommand{\hst}{{HST}\xspace}

\newcommand{\mrk}{{Mrk~509}\xspace}
\newcommand{\ngc}{{NGC~5548}\xspace}

\newcommand{\nustar}{{\it NuSTAR}\xspace}
\newcommand{\wise}{{{WO}}\xspace}

\newcommand{\ergcm}{{\ensuremath{\rm{erg\ cm}^{-2}\ \rm{s}^{-1}\ {\AA}^{-1}}}\xspace}
\newcommand{\ergflux}{{\ensuremath{\rm{erg\ cm}^{-2}\ \rm{s}^{-1}}}\xspace}

\begin{document}

\title{Anatomy of the AGN in NGC 5548}

\subtitle{I. A global model for the broadband spectral energy distribution}

\author{
M. Mehdipour \inst{1,2} 
\and
J.S. Kaastra \inst{1,3,4}
\and 
G.A. Kriss \inst{5,6}
\and
M. Cappi \inst{7}
\and
P.-O. Petrucci \inst{8,9}
\and
K.C. Steenbrugge \inst{10,11}
\and
N. Arav \inst{12}
\and
E. Behar \inst{13}
\and 
S. Bianchi \inst{14}
\and 
R. Boissay \inst{15}
\and
G. Branduardi-Raymont \inst{2}
\and
E. Costantini \inst{1}
\and
J. Ebrero \inst{16,1}
\and
L. Di Gesu \inst{1}
\and 
F.A. Harrison \inst{17}
\and
S. Kaspi \inst{13}
\and
B. De Marco \inst{18}
\and 
G. Matt \inst{14}
\and
S. Paltani \inst{15}
\and
B.M. Peterson \inst{19,20}
\and 
G. Ponti \inst{18}
\and
F. Pozo Nu\~{n}ez \inst{21}
\and
A. De Rosa \inst{22}
\and
F. Ursini \inst{8,9}
\and
C.P. de Vries \inst{1}
\and
D.J. Walton \inst{23,17}
\and
M. Whewell \inst{2}
}

\institute{
SRON Netherlands Institute for Space Research, Sorbonnelaan 2, 3584 CA Utrecht, the Netherlands\\ \email{M.Mehdipour@sron.nl}
\and
Mullard Space Science Laboratory, University College London, Holmbury St. Mary, Dorking, Surrey, RH5 6NT, UK
\and
Department of Physics and Astronomy, Universiteit Utrecht, P.O. Box 80000, 3508 TA Utrecht, the Netherlands
\and
Leiden Observatory, Leiden University, PO Box 9513, 2300 RA Leiden, the Netherlands
\and
Space Telescope Science Institute, 3700 San Martin Drive, Baltimore, MD 21218, USA
\and
Department of Physics and Astronomy, The Johns Hopkins University, Baltimore, MD 21218, USA
\and
INAF-IASF Bologna, Via Gobetti 101, I-40129 Bologna, Italy
\and
Univ. Grenoble Alpes, IPAG, F-38000 Grenoble, France
\and
CNRS, IPAG, F-38000 Grenoble, France
\and
Instituto de Astronom\'{i}a, Universidad Cat\'{o}lica del Norte, Avenida Angamos 0610, Casilla 1280, Antofagasta, Chile
\and
Department of Physics, University of Oxford, Keble Road, Oxford, OX1 3RH, UK
\and
Department of Physics, Virginia Tech, Blacksburg, VA 24061, USA
\and
Department of Physics, Technion-Israel Institute of Technology, 32000 Haifa, Israel
\and
Dipartimento di Matematica e Fisica, Universit\`{a} degli Studi Roma Tre, via della Vasca Navale 84, 00146 Roma, Italy
\and
Department of Astronomy, University of Geneva, 16 Ch. d'Ecogia, 1290 Versoix, Switzerland
\and
European Space Astronomy Centre, P.O. Box 78, E-28691 Villanueva de la Ca\~{n}ada, Madrid, Spain
\and
Cahill Center for Astronomy and Astrophysics, California Institute of Technology, Pasadena, CA 91125, USA
\and
Max-Planck-Institut f\"{u}r extraterrestrische Physik, Giessenbachstrasse, D-85748 Garching, Germany
\and
Department of Astronomy, The Ohio State University, 140 W 18th Avenue, Columbus, OH 43210, USA
\and
Center for Cosmology \& AstroParticle Physics, The Ohio State University, 191 West Woodruff Ave., Columbus, OH 43210, USA
\and
Astronomisches Institut, Ruhr-Universit\"{a}t Bochum, Universit\"{a}tsstra\ss e 150, 44801, Bochum, Germany
\and
INAF/IAPS - Via Fosso del Cavaliere 100, I-00133 Roma, Italy
\and
Jet Propulsion Laboratory, California Institute of Technology, 4800 Oak Grove Drive, Pasadena, CA 91109, USA
}

\date{Received 20 November 2014 / Accepted 18 December 2014}

\abstract
{
An extensive multi-satellite campaign on NGC~5548 has revealed this archetypal Seyfert-1 galaxy to be in an exceptional state of persistent heavy absorption. Our observations taken in 2013--2014 with \xmm, \swift, \nustar, \integral, \chandra, HST and two ground-based observatories have together enabled us to establish that this unexpected phenomenon is caused by an outflowing stream of weakly ionised gas (called the obscurer), extending from the vicinity of the accretion disk to the broad-line region. In this work we present the details of our campaign and the data obtained by all the observatories. We determine the spectral energy distribution of \ngc from near-infrared to hard X-rays by establishing the contribution of various emission and absorption processes taking place along our line of sight towards the central engine. We thus uncover the intrinsic emission and produce a broadband continuum model for both obscured (average summer 2013 data) and unobscured ($<$ 2011) epochs of \ngc. Our results suggest that the intrinsic NIR/optical/UV continuum is a single Comptonised component with its higher energy tail creating the `soft X-ray excess'. This component is compatible with emission from a warm, optically-thick corona as part of the inner accretion disk. We then investigate the effects of the continuum on the ionisation balance and thermal stability of photoionised gas for unobscured and obscured epochs. 
}

\authorrunning{M. Mehdipour et al.}
\titlerunning{Anatomy of the AGN in NGC 5548. I.}
\maketitle

\section{Introduction}

The supermassive black holes (SMBHs) at the heart of active galactic nuclei (AGN) grow through accretion of matter from their host galaxies. This accretion is accompanied by outflows (powerful relativistic jets of plasma and/or winds of ionised gas), which transport matter and energy away from the nucleus, thus linking the SMBHs to their host galaxies. Understanding the feedback mechanisms between SMBHs and their environments is important in AGN science, as well as in cosmology. They can have significant impacts and implications for the evolution of SMBHs and star formation in their host galaxies (e.g. \citealt{Silk98,King10}), chemical enrichment of their surrounding intergalactic medium (e.g. \citealt{Oppen06}), and the cooling flows at the core of galaxy clusters (e.g. \citealt{Ciott01}).
\\

The ionised outflows are an essential component of the energy balance of AGN and are best studied through high-resolution X-ray and UV spectroscopy of their absorption line spectra. The Reflection Grating Spectrometer (RGS) of \xmm and the Low-Energy \& High-Energy Transmission Grating Spectrometers (LETGS and HETGS) of \chandra, together with the Cosmic Origins Spectrograph (COS) of the Hubble Space Telescope (HST) have vastly advanced our knowledge of these outflows in recent years (see e.g. the review by \citealt{Costan10}). With current instrumentation, the nearby and bright Seyfert type AGN are the best laboratories for studying these outflows. The majority of the absorbing gas is detectable in the X-ray band. The outflows detected in the soft X-rays (referred to as `warm absorbers') are found to have column densities of $10^{20}$$-$$10^{24}$~cm$^{-2}$ and consist of multiple phases of photoionised gas with temperatures of $10^{4}$$-$$10^{6}$~K, travelling at velocities of up to a few $10^3$~$\rm{km}\ \rm{s}^{-1}$ (see e.g. \citealt{Blu05}). Understanding the origin and launching mechanism of the ionised outflows in AGN is an active area of research. It has been suggested that these outflows are produced by irradiation of the dusty gas torus structure, which surrounds the SMBH and accretion disk (e.g. \citealt{Krol01}), or that they originate as radiatively driven winds from the accretion disk (e.g. \citealt{Bege83,Prog00}), or are in the form of magnetohydrodynamic (MHD) winds from the accretion disk (e.g. \citealt{Koni94,Bott00}).
\\

Many aspects and physical properties of these outflows are still poorly understood. For instance, the location of the ionised absorbers needs to be established to distinguish between the different outflow mechanisms and to determine their mass outflow rates and kinetic luminosities, which are essential parameters in assessing their impact on their surroundings and their contribution to AGN feedback (e.g. \citealt{Hopk10}). The distance of an X-ray or UV absorbing outflow to the ionising source can be determined from estimates of the density of the absorbing gas. The density can be measured from either density-sensitive UV lines (e.g. \citealt{Gab05}) or from the recombination timescale of the ionised absorber \citep{Kaas12}. In the X-ray band, the latter delivers more robust density measurements than the use of density-sensitive X-ray lines. In the recombination timescale method, as the intrinsic luminosity of the AGN varies over time, the ionisation state of the absorber changes with a time delay; by measuring this lag, the electron density and hence the distance of the absorber to the ionising source can be obtained. Such a study was first done as part of our 2009 multi-wavelength campaign on the Seyfert-1/QSO \object{Mrk~509} \citep{Kaas12}. From the response of three absorber components with the highest ionisation in the soft X-rays, individual distances of 5--100~pc were derived, pointing to an origin in the narrow-line region (NLR) or the AGN torus region.
\\

The archetypal Seyfert-1 galaxy \object{NGC 5548} is one of the most widely studied nearby active galaxies. It was one of the 12 classified objects in the seminal work of \citet{Seyf43}, and since the sixties it has been the target of various AGN studies. In more recent times it was the first object in which narrow X-ray absorption lines from warm absorbers were discovered by \citet{Kaa00} using a high-resolution \chandra LETGS spectrum. Later on, a detailed study of its warm absorber was carried out by \citet{Ste05}. Furthermore, extensive optical/UV reverberation mapping studies of \ngc (e.g. \citealt{Pete02,Panc13}) have provided one of the most detailed pictures available of the broad-line region (BLR) size and structure in this AGN. Prior to our recent campaign on \ngc, the X-ray properties and variability of \ngc were known to be typical of standard Seyfert-1 AGN sharing common characteristics (e.g. \citealt{Bian09,Pont12}).
\\

In 2013--2014 we carried out an ambitious multi-wavelength campaign on \ngc, which was similar but more extensive than our campaign on \mrk. This remarkable campaign has utilised eight observatories to take simultaneous and frequent observations of the AGN. It incorporates instruments onboard five X-ray observatories: \xmm \citep{Jans01}, \swift \citep{Gehr04}, \nustar \citep{Harr13}, \integral \citep{Wink03}, \chandra's LETGS \citep{Brink00}, as well as the HST COS \citep{Green12}, and two ground-based optical observatories: the {\it Wise Observatory} (WO) and the Observatorio Cerro Armazones (OCA). These observatories have collected over 2.4 Ms of X-ray and 800 ks of optical/UV observation time. As previously reported by \citet{Kaas14}, \ngc was discovered to be obscured in X-rays with mainly narrow emission features imprinted on a heavily absorbed continuum. This obscuration is thought to be caused by a stream of clumpy weakly-ionised gas located at distances of about 2--7 light days from the black hole and partially covering the X-ray source and the BLR. From its associated broad UV absorption lines detected in HST COS spectra, the obscurer is found to be outflowing with velocities of up to 5000 \kms. The intense \swift monitoring on \ngc shows the obscuration has been continuously present for a few years (at least since Feb 2012). As the ionising UV/X-ray radiation is being shielded by the obscurer, new weakly-ionised features of UV and X-ray absorber outflows have been detected. Compared to normal warm absorber outflows commonly seen in Seyfert-1s at pc scale distances, the remarkable obscurer in \ngc is a new breed of weakly-ionised, higher-velocity outflowing gas, which is much closer to the black hole and extends to the BLR. As reported in \citet{Kaas14} the outflowing obscurer is likely to originate from the accretion disk. Based on the high outflow velocity of the obscurer, its short-timescale absorption variability, and its covering fractions of the continuum and the BLR, the obscurer is in close proximity to the central source, and its geometry extends from near the disk to outside the BLR.
\\

In this work we present a broadband spectral analysis of the \ngc data. The structure of the paper is as follows. Section \ref{campaign_sect} gives an overview of our multi-satellite campaign. In Sect. \ref{lc_sect} we present lightcurves of \ngc constructed at various energies from near-infrared (NIR) to hard X-rays. In Sect. \ref{data_correct_sect} we explain the required steps in determining the spectral energy distribution (SED) of \ngc. In Sect. \ref{soft_excess_sect} we examine the soft X-ray excess in \ngc and present an appropriate model for it. In Sec. \ref{broadband_sect} we describe the modelling of the broadband continuum in unobscured and obscured epochs and present our results. The thermal stability curves, corresponding to various ionising SEDs, are presented in Sect. \ref{stability_sect}. We discuss all our findings in Sect. \ref{discussion} and give concluding remarks in Sect. \ref{conclusions}. The processing of the data from all the instruments is described in Appendix \ref{data_appendix}.
\\

The spectral analysis and modelling, presented in this work, were done using the {\tt SPEX}\,\footnote{\url{http://www.sron.nl/spex}} package \citep{Kaa96} version 2.05.02. We also made use of tools in NASA's {\tt HEASOFT}\,\footnote{\url{http://heasarc.nasa.gov/lheasoft}} v6.14 package. The spectra shown in this report are background-subtracted and are displayed in the observed frame, unless otherwise stated in the text. We use C-statistics \citep{Cash79} for spectral fitting (unless otherwise stated) and give errors at $1\sigma$ (68\%) confidence level. The redshift of \ngc is set to 0.017175 \citep{deVa91} as given in the NASA/IPAC Extragalactic Database (NED). The adopted cosmological parameters for luminosity computations in our modelling are ${H_{0}=70\ \mathrm{km\ s^{-1}\ Mpc^{-1}}}$, $\Omega_{\Lambda}=0.70$ and $\Omega_{m}=0.30$.

\section{Multi-wavelength campaign on \ngc}
\label{campaign_sect}

At the core of our campaign in summer 2013 (22 June to 1 August), there were 12 \xmm observations, of which five were taken simultaneously with HST COS, four with \integral and two with \nustar observations. Throughout our campaign and beyond, \swift monitored NGC 5548 on a daily basis. There were also optical monitorings with \wise and OCA. The summer \xmm observations were followed by three \chandra LETGS observations taken in the first half of September 2013, one of which was taken simultaneously with \nustar. The September observations were triggered upon observing a large jump in the X-ray flux from our \swift monitoring. However, due to scheduling constraints by the time the triggered \chandra observations were made, the week-long peak of high X-ray flux was just missed and the tail end of the flare was caught. Nonetheless, the X-ray flux was still higher than during the \xmm observations and improved LETGS spectra were obtained. During this autumn period (Sep-Nov 2013), \ngc was not visible to \xmm and thus no \xmm observations could be triggered. 
\\

A few months later, two more \xmm observations were taken, one in December 2013 and the other in February 2014. The former observation was simultaneous with HST COS and \nustar observations, and the latter close in time to an INTEGRAL observation. The timeline of all the observations in our campaign is displayed in Fig. \ref{timeline_fig} and the observation logs are provided in Table \ref{obslog}. In Appendix \ref{data_appendix}, we describe the observations made by the instruments of each observatory and give details on their data reduction and processing.

%
\begin{figure*}[!]
\centering
\resizebox{0.94\hsize}{!}{\includegraphics[angle=0]{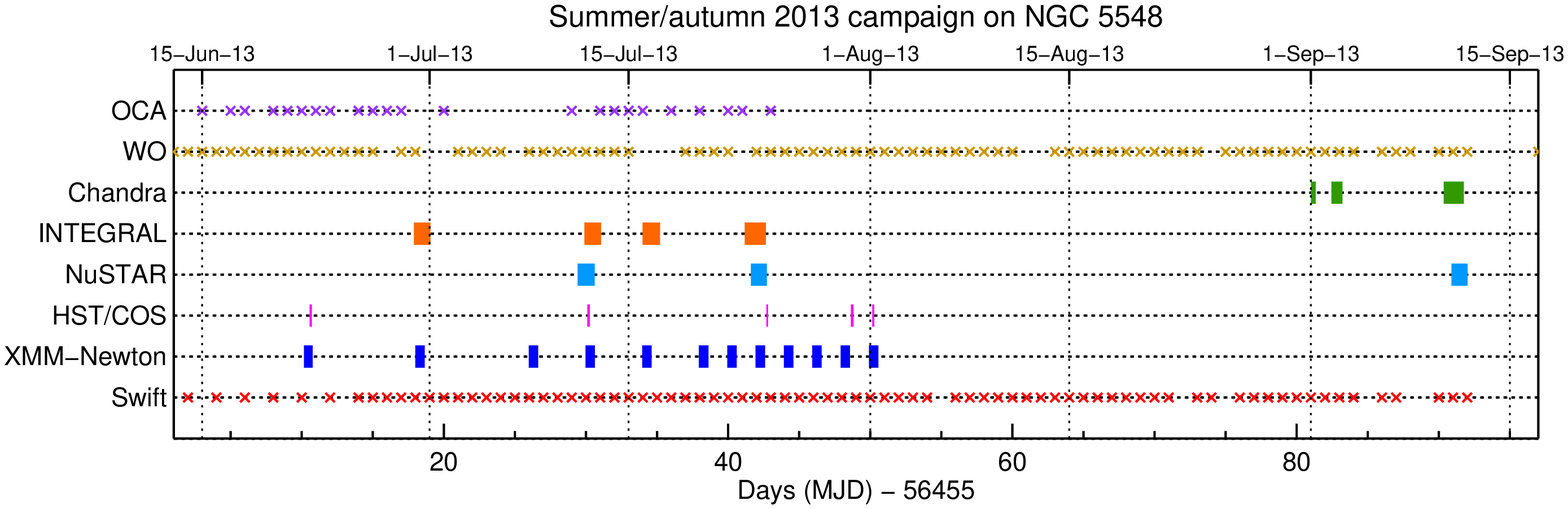}}\vspace{-0.6cm}
\resizebox{0.94\hsize}{!}{\includegraphics[angle=0]{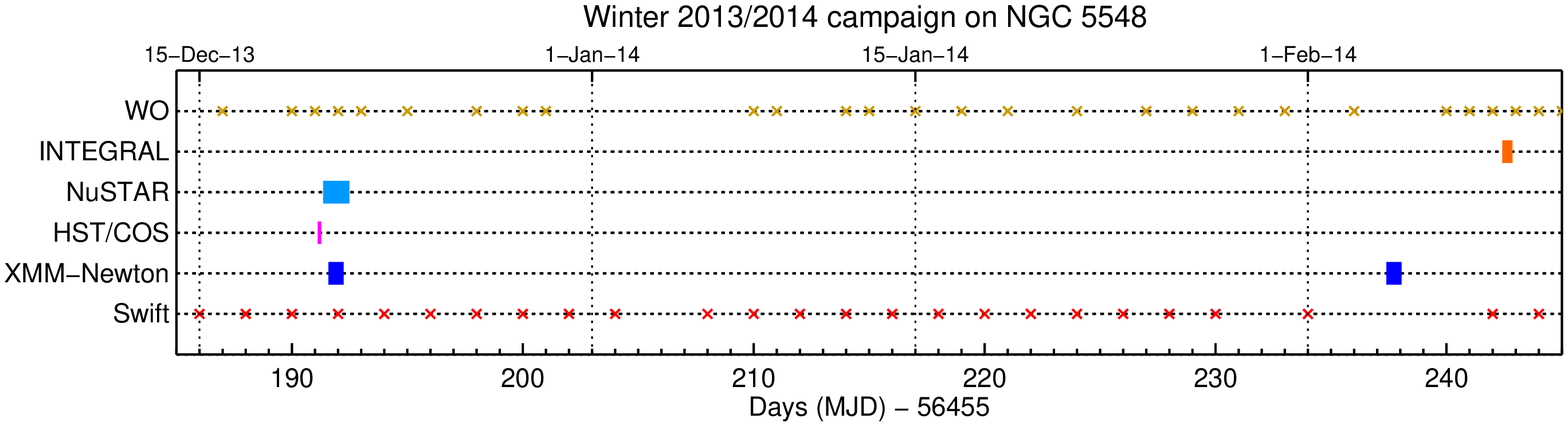}}\vspace{-0.35cm}
\caption{Timeline of our multi-wavelength campaign on \ngc. The thickness of each rectangular symbol on the time axis is indicative of the length of that observation. The days in which \swift, \wise and OCA made observations are indicated by crosses.}
\label{timeline_fig}
\end{figure*}

%
\begin{table*}[!]
\begin{minipage}[t]{\hsize}
\setlength{\extrarowheight}{3pt}
\caption{Observation log of the \ngc campaign. For the \swift, OCA and WO monitorings, the Obs. numbers correspond to days in which observations were taken. For \swift, only recent observations taken in 2013--2014 are reported, including all monitoring programs during this period. The \swift lengths in ks are the total length of the observations in each year. For HST COS the span of each observation in ks is given.}
\label{obslog}
\centering
\renewcommand{\footnoterule}{}
\tiny
\begin{tabular}{l c c c c || l c c c c}
\hline \hline
 & & & \multicolumn{1}{c}{Start time (UTC)} & Length &  & & & \multicolumn{1}{c}{Start time (UTC)} & Length  \\
Observatory & Obs. & ID & yyyy-mm-dd\ \ hh:mm & (ks) & Observatory & Obs. & ID & yyyy-mm-dd\ \ hh:mm & (ks) \\ 
\hline

\xmm  & 1 & 0720110301 		& 2013-06-22\ \ 03:53	& 50.5			& HST COS & 1 & lc7001 & 2013-06-22\ \ 13:25 & 13.0 \\
\xmm  & 2 & 0720110401		& 2013-06-29\ \ 23:50	& 55.5			& HST COS & 2 & lc7002 & 2013-07-12\ \ 02:23 & 14.2 \\
\xmm  & 3 & 0720110501		& 2013-07-07\ \ 23:28	& 50.9			& HST COS & 3 & lc7003 & 2013-07-24\ \ 16:43 & 8.9 \\
\xmm  & 4 & 0720110601		& 2013-07-11\ \ 23:11	& 55.5			& HST COS & 4 & lc7004 & 2013-07-30\ \ 15:15 & 16.0 \\
\xmm  & 5 & 0720110701		& 2013-07-15\ \ 22:56	& 55.5			& HST COS & 5 & lc7005 & 2013-08-01\ \ 03:21 & 12.3 \\
\xmm  & 6 & 0720110801		& 2013-07-19\ \ 22:40	& 56.5			& HST COS & 6 & lc7006 & 2013-12-20\ \ 02:49 & 13.0 \\ \cline{6-10}
\xmm  & 7 & 0720110901		& 2013-07-21\ \ 22:32	& 55.5			& \nustar & 1 & 60002044002 & 2013-07-11\ \ 09:50 & 51.6 \\
\xmm  & 8 & 0720111001		& 2013-07-23\ \ 22:24	& 55.5			& & & 60002044003 & 2013-07-12\ \ 00:10 & 52.2 \\
\xmm  & 9 & 0720111101		& 2013-07-25\ \ 22:15	& 55.5			& \nustar & 2 & 60002044005 & 2013-07-23\ \ 14:25 & 97.2 \\
\xmm  & 10 & 0720111201	& 2013-07-27\ \ 22:06	& 55.5			& \nustar & 3 & 60002044006 & 2013-09-10\ \ 21:25 & 97.5 \\
\xmm  & 11 & 0720111301		& 2013-07-29\ \ 21:58	& 50.4			& \nustar & 4 & 60002044008 & 2013-12-20\ \ 08:30 & 98.1 \\ \cline{6-10}
\xmm  & 12 & 0720111401	& 2013-07-31\ \ 21:49	& 55.5			& \chandra & 1 & 16369 & 2013-09-01\ \ 00:01 & 29.7 \\
\xmm  & 13 & 0720111501	& 2013-12-20\ \ 14:01	& 55.3			& \chandra & 2 & 16368 & 2013-09-02\ \ 10:33 & 67.5 \\
\xmm  & 14 & 0720111601	& 2014-02-04\ \ 09:33	& 55.5			& \chandra & 3 & 16314 & 2013-09-10\ \ 08:17 & 122.0 \\ \cline{6-10}
\cline{1-5}
\integral  & 1 & 10700010001	& 2013-06-29\ \ 21:34 	& 100.0			& \swift (2013)\,\footnote{The \swift monitoring in 2013 ended on 2013-12-31 18:19.} & 1--160 & \footnote{The \swift target IDs of \ngc in 2013: 30022, 80131, 91404, 91711, 91737, 91739, 91744, 91964.} & 2013-01-04\ \ 00:24 & 326.6 \\
\integral  & 2 & 10700010002	& 2013-07-11\ \ 21:13 	& 102.0			& \swift (2014)\,\footnote{The \swift monitoring up to 2014-07-01 00:00 is reported here. The \swift monitoring of \ngc is currently ongoing in 2014--2015.}  & 161--291 & \footnote{The \swift target IDs of \ngc in 2014: 30022, 33204, 91964.} & 2014-01-02\ \ 14:53 & 182.5 \\
\integral  & 3 & 10700010003	& 2013-07-15\ \ 23:31 	& 106.5			& OCA\,\footnote{The OCA monitoring ended on 2013-07-25 01:13. Observations taken in the B, V, R filters, with 150~s exposure in each filter.} & 1--27 & - & 2013-05-20\ \ 03:17 & $^e$ \\ 
\integral  & 4 & 10700010004 	& 2013-07-23\ \ 03:54 	& 128.9			& WO (2013)\,\footnote{The \wise monitoring ended on 2013-09-24 04:59. Observations taken in the B, V, R, I filters, with 300~s exposure in each filter.} & 1--93 & - & 2013-06-02\ \ 07:22 & $^f$ \\
\integral  & 5 & 11200110003 	& 2014-02-09\ \ 10:00 	& 38.2			& WO (2014)\,\footnote{The \wise monitoring ended on 2014-04-14 08:25. Observations taken in the B, V, R, I filters, with 300~s exposure in each filter.} & 94--150 & - & 2013-12-16\ \ 14:38 & $^g$ \\ 

\hline
\end{tabular}
\end{minipage}
\end{table*}

\section{Lightcurves of \ngc}
\label{lc_sect}

Here we present lightcurves of \ngc, from NIR to hard X-rays, obtained from the eight observatories used during our summer 2013 campaign. 

%
\begin{figure*}[!]
\centering
\resizebox{0.742\hsize}{!}{\includegraphics[angle=0]{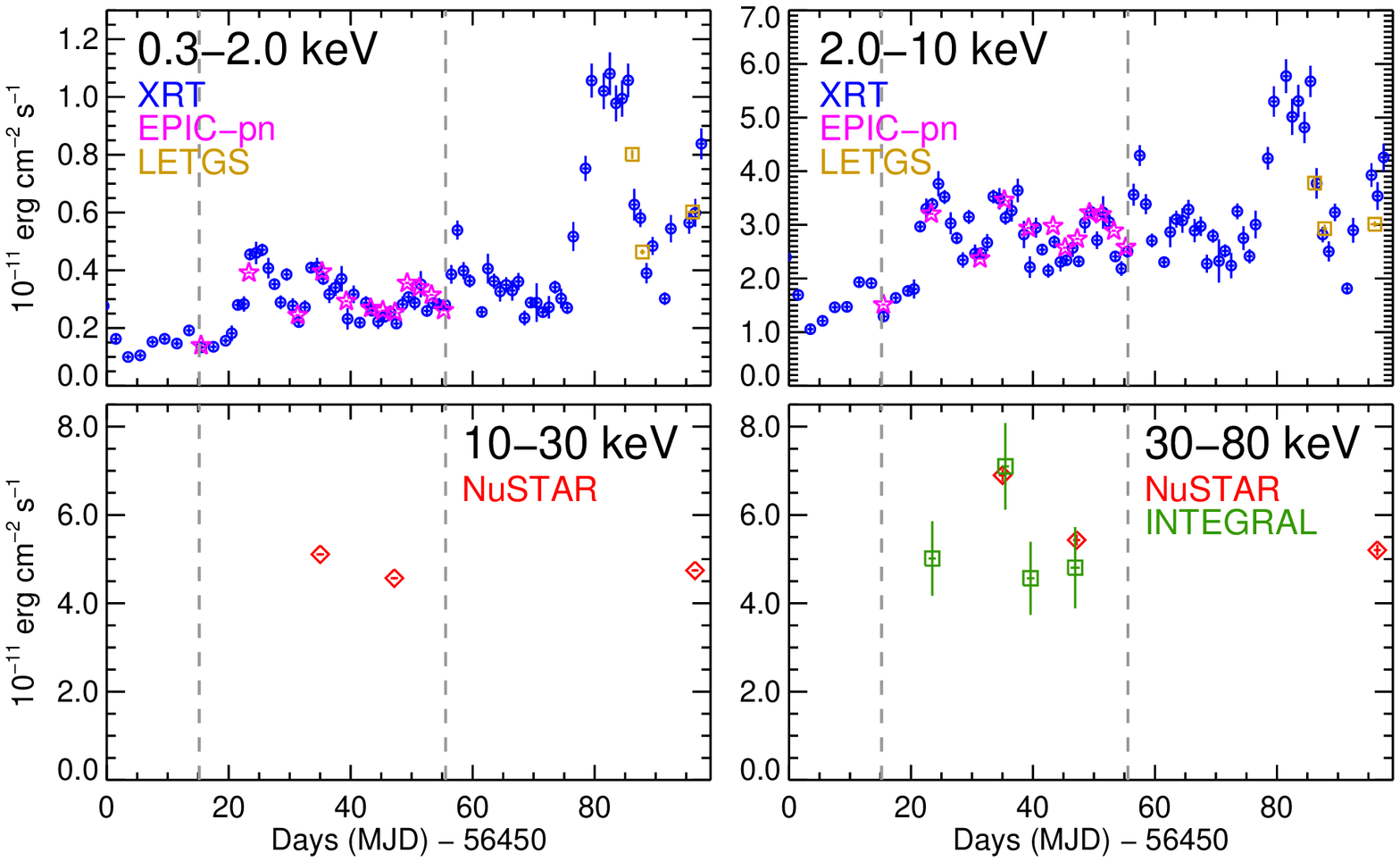}}
\vspace{-0.1cm}
\caption{X-ray lightcurves of \ngc from our summer 2013 campaign. The lightcurves show the observed flux and are displayed between 7 June and 14 Sep 2013. The XRT data are shown as blue circles, EPIC-pn as magenta stars, LETGS as dark yellow squares, \nustar as red diamonds and INTEGRAL data as green squares. The vertical dashed lines indicate the interval between \xmm Obs. 1 and 12.}
\label{xray_lc}
%
\centering
\resizebox{0.711\hsize}{!}{\includegraphics[angle=0]{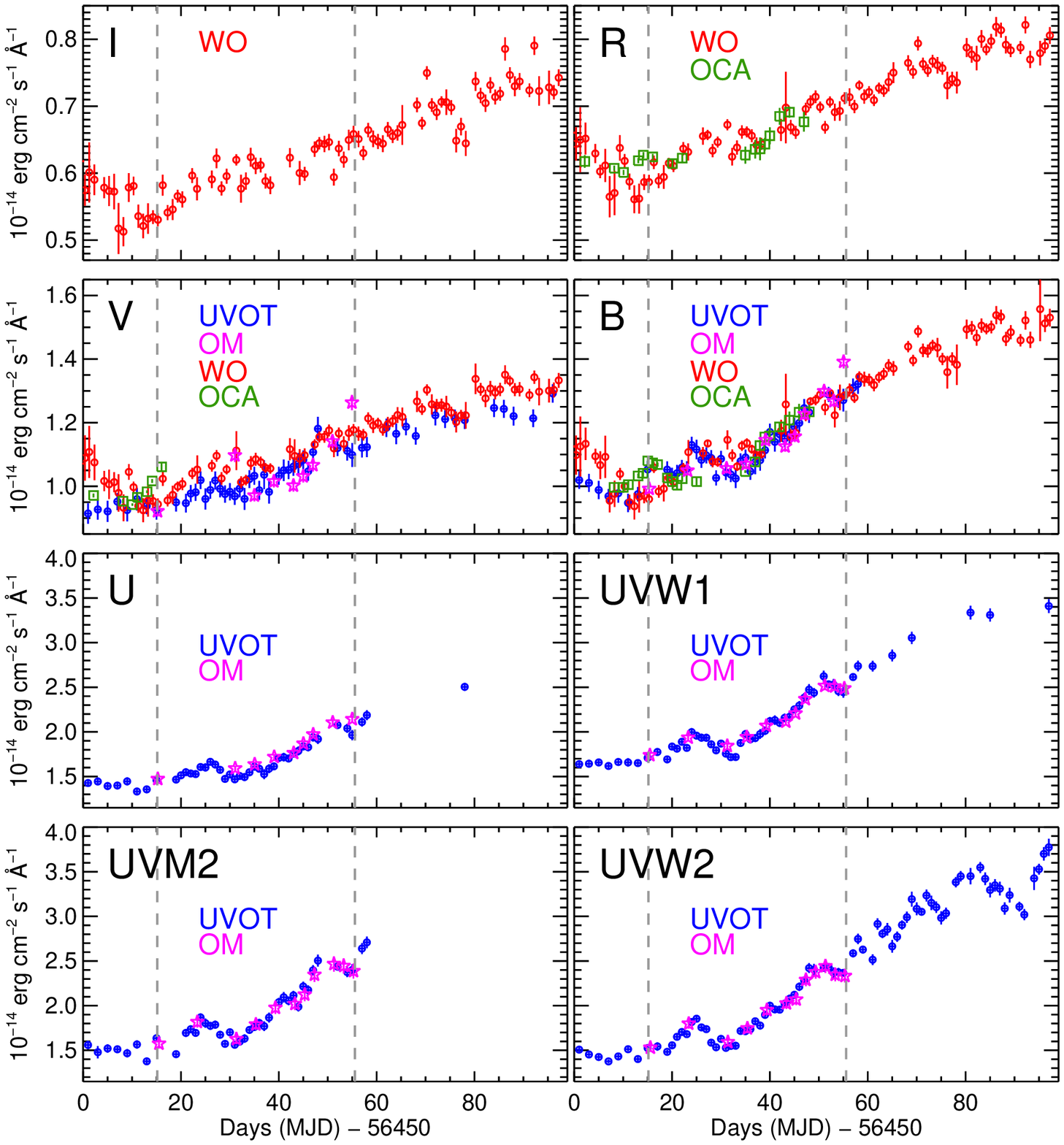}}
\vspace{-0.1cm}
\caption{NIR, optical and UV lightcurves of \ngc taken with various filters from our summer 2013 campaign. The lightcurves show the observed energy flux and are displayed between 7 June and 14 Sep 2013. The UVOT data are shown as blue circles, OM as magenta stars, \wise as red circles, and OCA data as green squares. The vertical dashed lines indicate the interval between \xmm Obs. 1 and 12.}
\label{optical_lc}
\end{figure*}

\subsection{X-ray lightcurves}

Figure \ref{xray_lc} shows the X-ray lightcurves of \ngc, from the \swift, \xmm, \nustar, \integral and \chandra observations, with the average energy flux of each observation calculated over four X-ray energy bands from soft to hard X-rays. The X-ray fluxes have been calculated by fitting each dataset over the required energy band with the model described later on in Sect. \ref{broadband_sect}.
\\

For the \xmm observations, the lowest X-ray fluxes were observed in Obs. 1: ${F_{{\rm 0.3-2\ keV}}} = 1.40 \times 10^{-12}\ \rm{erg\ cm}^{-2}\ \rm{s}^{-1}$ and ${F_{{\rm 2-10\ keV}}} = 1.51 \times 10^{-11}\ \rm{erg\ cm}^{-2}\ \rm{s}^{-1}$. These fluxes, compared to those from the unobscured 2000 and 2001 observations, are smaller by factors ranging from 17--27 (for 0.3--2 keV) and 2--3 (for 2--10 keV). Later on, in the summer \xmm campaign, the X-ray fluxes increased from their minimum in Obs. 1 by a factor of 2.8 (0.3--2 keV) and 2.3 (2--10 keV), peaking at \xmm Obs. 4. The X-ray flux during the summer \xmm campaign was low relative to other obscured epochs, as seen from the long-term \swift monitoring. The average fluxes of the summer 2013 \xmm observations (${F_{{\rm 0.3-2\ keV}}} = 3.02 \times 10^{-12}\ \rm{erg\ cm}^{-2}\ \rm{s}^{-1}$ and ${F_{{\rm 2-10\ keV}}} = 2.77 \times 10^{-11}\ \rm{erg\ cm}^{-2}\ \rm{s}^{-1}$) are smaller by a factor of 1.6 (for 0.3--2 keV) and 1.1 (for 2--10 keV) than the average fluxes for the whole obscured epoch observed with \swift between February 2012 and July 2014 (\citealt{Meh14b}).
\\

After the end of the summer \xmm observations (Obs. 1--12), \swift kept on monitoring \ngc until the middle of September 2013, when it passed out of \swift's visibility window. However, prior to that a sudden increase in the X-ray flux was observed with \swift which started on 24 August 2013 (56528 in MJD) and lasted about one week. Compared to the average flux of the summer \xmm observations, the flux increased by a factor of 3.4 (for 0.3--2 keV) and 1.9 (for 2--10 keV). This resulted in the triggering of the \chandra LETGS observations. Later on, the two final \xmm observations (Obs. 13 and 14) were taken in December 2013 and February 2014 (these are outside the range of Fig. \ref{xray_lc} lightcurves). Compared to the average \xmm flux during the summer 2013 observations, the Obs. 13 flux was slightly lower by a factor of 1.2 (for 0.3--2 keV) and 1.1 (for 2--10 keV), and for Obs. 14 the flux was higher by a factor of 1.5 for 0.3--2 keV and about the same for 2--10 keV. A comprehensive analysis of the \xmm lightcurves and their properties is reported in \citet{Cappi14}.

\subsection{NIR/optical/UV lightcurves}
%

%
\begin{figure*}[!]
\centering
\vspace{-0.2cm}
\resizebox{0.9\hsize}{!}{\hspace{-0.7cm}\includegraphics[angle=0]{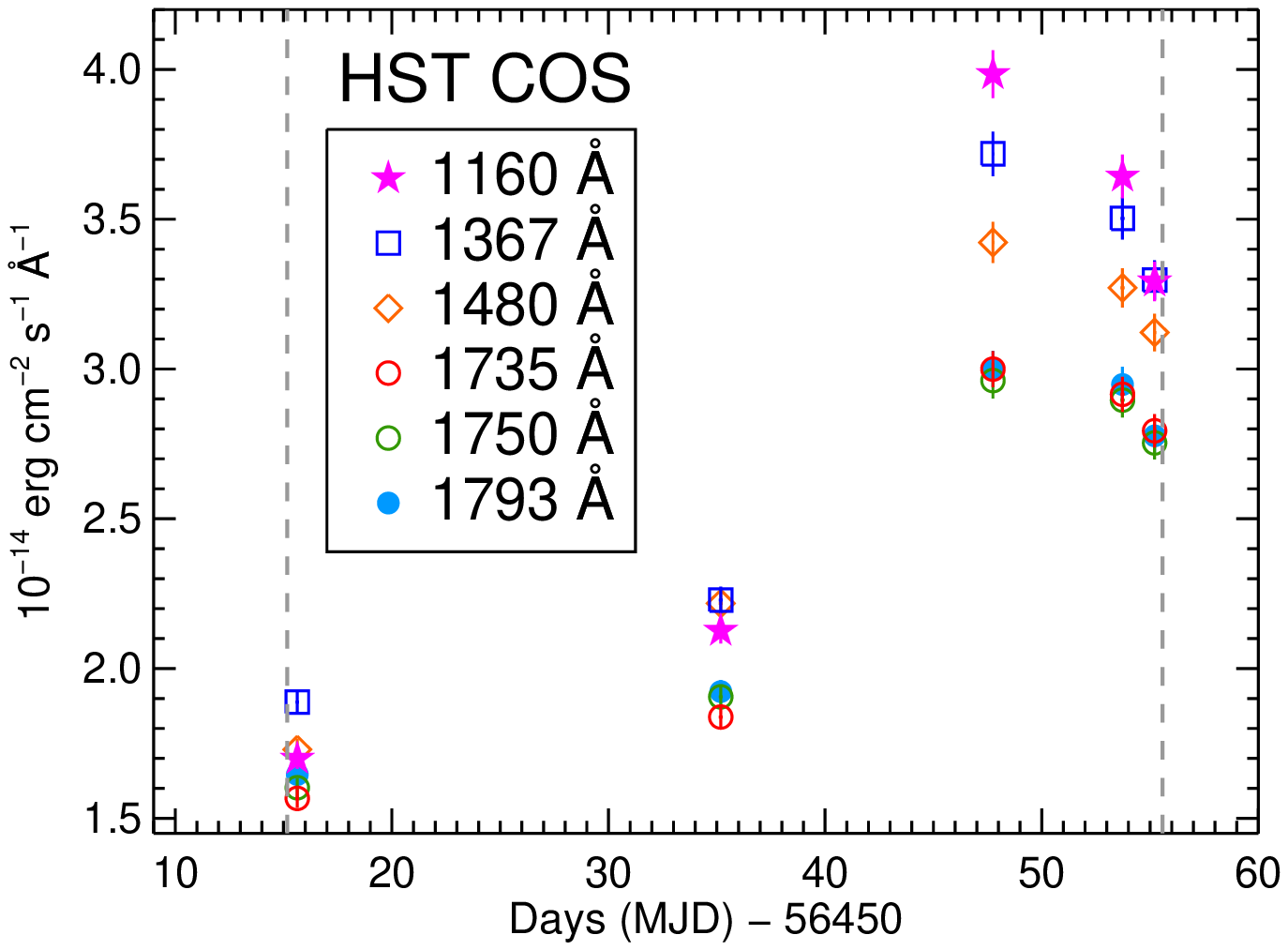}\hspace{-0.7cm}\includegraphics[angle=0]{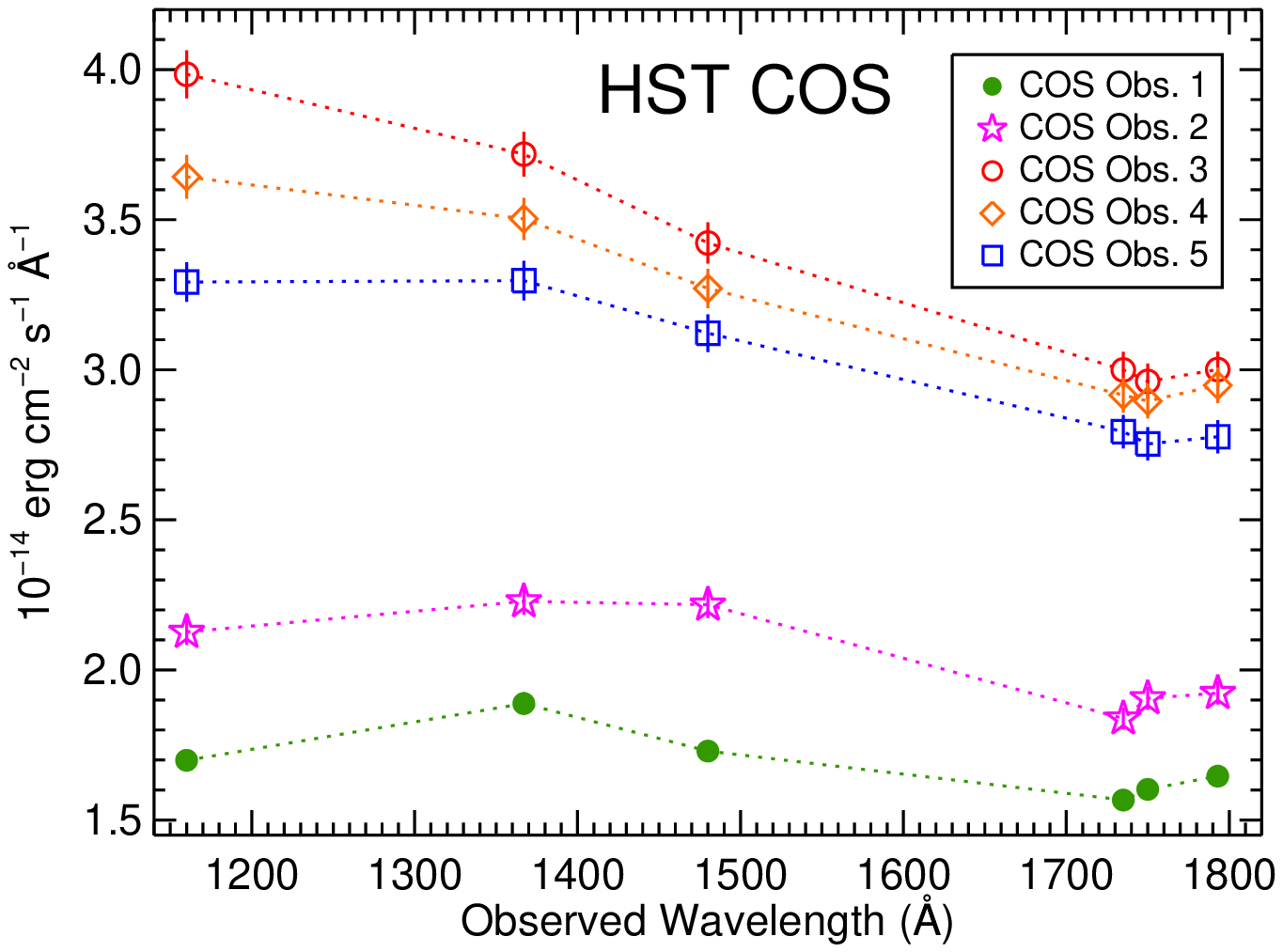}}
\caption{HST COS fluxes of \ngc at different UV wavelengths from summer 2013 observations. The lightcurves ({\it left panel}) and spectra ({\it right panel}) show the observed energy flux obtained from narrow energy bands which are free of spectral features. The displayed range of the lightcurves is between 16 June and 6 August 2013, with the vertical dashed lines indicating the interval between \xmm Obs. 1 and 12.}
\label{cos_lc}
\end{figure*}

In Fig. \ref{optical_lc} we show the optical/UV lightcurves from \swift UVOT, \xmm OM, \wise and OCA, taken in the I, R, V, B, U, UVW1, UVM2 and UVW2 filters. Figure \ref{cos_lc} shows the HST COS fluxes variability at six narrow UV energy bands (which are free of spectral features and represented by their central wavelengths) between 1160 and 1793 \AA\ for the summer 2013 observations. To calculate the flux in broadband photometric filters, one requires convolution of the response of the instrument and the transmission of the filter as a function of wavelength, with the shape of the optical/UV spectrum, to convert from instrumental units to energy flux at an effective wavelength. For the shape of the optical/UV spectrum, rather than using default spectral shapes commonly incorporated in the data reduction softwares (such as standard stars spectra, or a power-law), we used the optical/UV spectrum model that we derived for \ngc from modelling the OM and UVOT Optical and UV grism spectra and the HST COS continuum data (see Sect. \ref{data_correct_sect} and Fig. \ref{grism_fig}). This approach allows for a more accurate flux calculation in the photometric filters, taking into account the presence of strong AGN emission lines with a realistic continuum model. For instance, for the six \swift UVOT filters, the custom-fit spectral model results in 5 to 15\% flux improvement in our knowledge of the continuum flux in each filter. The calculation of the COS fluxes at the narrow UV energy bands is described in Appendix \ref{cos_sect}. In the lightcurves of Fig. \ref{optical_lc}, the UVOT and OM fluxes have been calculated at the following wavelengths for each filter: V (5402 \AA), B (4329 \AA), U (3501 \AA), UVW1 (2634 \AA), UVM2 (2231 \AA), UVW2 (2030 \AA). For OCA and \wise the fluxes have been given at the following wavelengths for each filter: I (8060 \AA), R (7000 \AA), V (5500 \AA), B (4330 \AA). 
\\

As illustrated in the Fig. \ref{optical_lc} lightcurves, the NIR/optical/UV fluxes were more or less continuously increasing throughout our summer \xmm campaign (Obs. 1--12) and beyond, until the end of our \swift monitoring in middle of September 2013, when the source was not visible to \swift for a couple of months. In fact the optical/UV fluxes recorded by \swift at mid September were the highest observed in all of the \swift monitoring of \ngc spanning from 2005 to 2014, which also coincided with our triggered \chandra LETGS observations. Figure \ref{cos_lc} shows that the COS UV fluxes increased from their minimum at \xmm Obs.1 (COS Obs. 1) and peaked at \xmm Obs. 8 (COS Obs. 3), followed by a decrease afterwards. The COS flux change from the first observation gets larger towards higher UV energies: factor of 1.8 at 1793 \AA\ to factor of 2.3 at 1160 \AA. 

\section{Determination of the continuum emission}
\label{data_correct_sect}

In order to establish the actual intrinsic continuum in the NIR/optical/UV part of the SED, the following absorption effects and emission contributions must be taken into account: (1) Galactic interstellar reddening; (2) Host galaxy stellar emission; (3) Emission lines from the broad-line region (BLR) and narrow-line region (NLR); (4) \ion{Fe}{ii} blended emission feature; (5) Balmer continuum emission. Moreover, in the X-ray band the following must be taken into account: (6) Galactic X-ray absorption; (7) Absorption by the obscurer; (8) Absorption by the warm absorber; (9) Soft X-ray emission lines.
\\

\begin{figure*}[!]
\centering
\vspace{-0.2cm}
\resizebox{1.00\hsize}{!}{\hspace{-0.9cm}\includegraphics[angle=0]{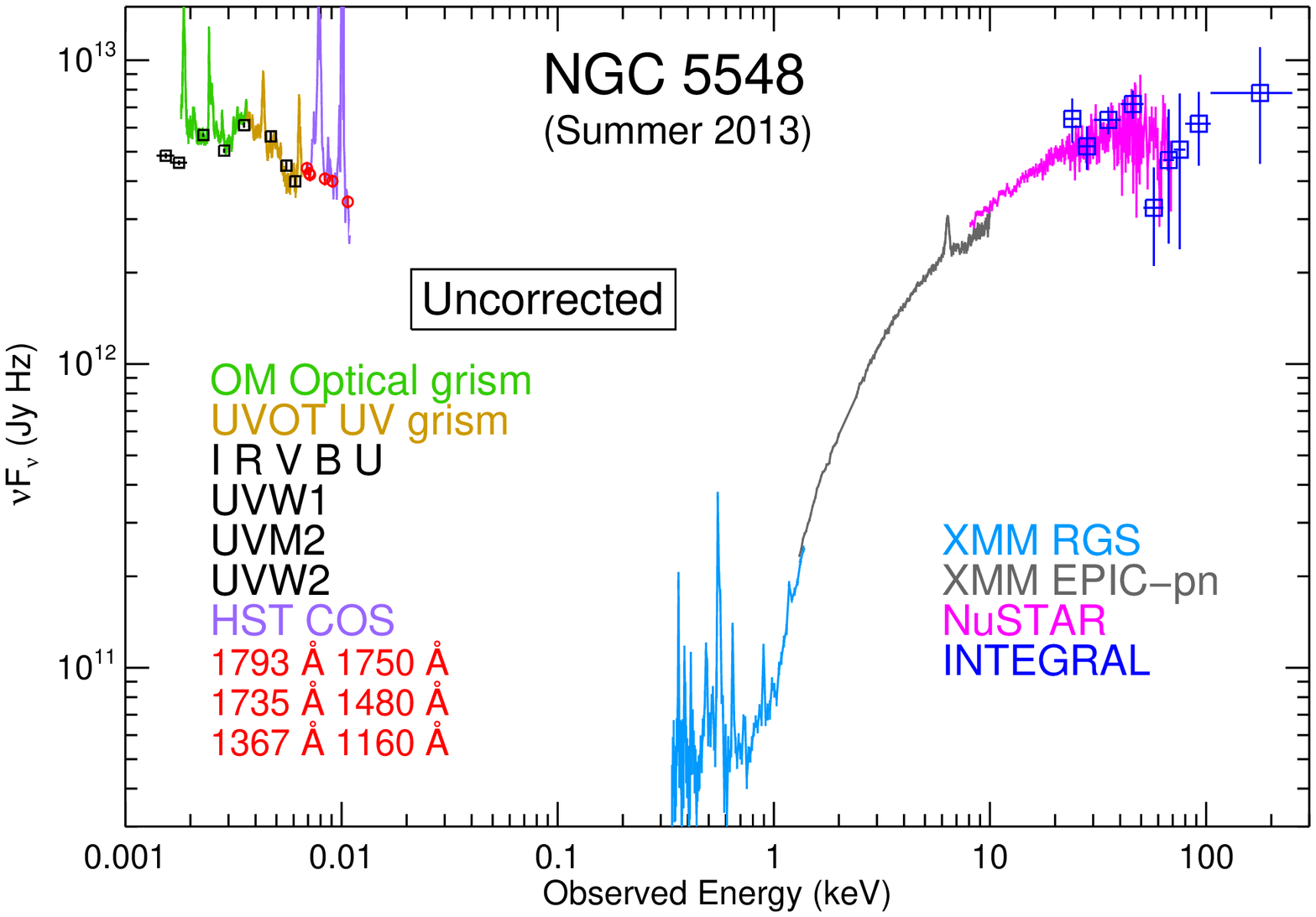}\hspace{-2.4cm}\includegraphics[angle=0]{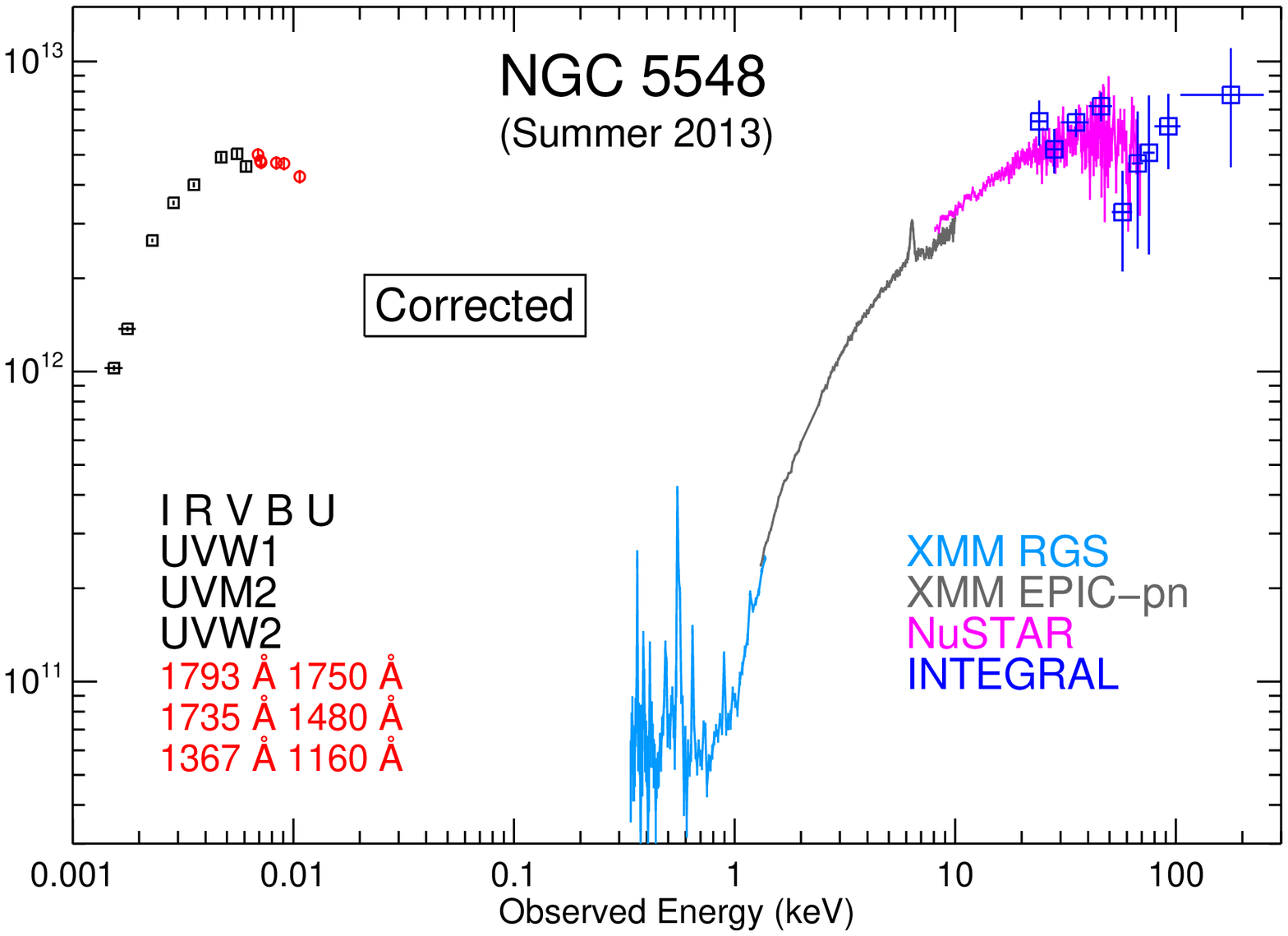}}\vspace{-0.1cm}
\caption{Average multi-wavelength data of NGC 5548 from our summer 2013 campaign. The data are displayed before ({\it left panel}) and after ({\it right panel}) corrections for Galactic reddening (Sect \ref{reddening_sec}), host galaxy stellar emission (Sect \ref{host_gal_sect}), emission lines from the BLR and NLR (Sect \ref{BLR_NLR_sect}), blended \ion{Fe}{ii} emission (Sect. \ref{FeII_sect}), Balmer continuum (Sect. \ref{Balmer_sect}) and Galactic interstellar X-ray absorption (Sect \ref{Gal_abs_sect}). The spectra have been binned for clarity of presentation. The displayed X-ray data include the effects of heavy absorption by the obscurer and the warm absorber. The shape of the NIR/optical/UV data after the corrections ({\it right panel}) is remarkable and exhibits a thermal disk spectrum.}
\label{obs_sed_fig}
\end{figure*}

In the following subsections, we describe the modelling of the above NIR/optical/UV components (1--5) in Sects. \ref{reddening_sec}--\ref{Balmer_sect} and the modelling of the X-ray components (6--9) in Sects. \ref{Gal_abs_sect}--\ref{xray_emission_line_sect}. We have modelled the grism spectra from OM and UVOT together with simultaneous HST COS continuum measurements to establish the contribution of each component and consequently correct the data taken in the photometric filters. For fitting the grism spectra in {\tt SPEX} we used $\chi^2$ minimisation instead of C-statistics (used for X-ray spectra), as the optical/UV spectra have sufficient counts per bin. In Fig. \ref{grism_fig} the best-fit model to the stacked OM and UVOT grism spectra and HST COS continua obtained from our simultaneous 2013 observations are shown. In Fig. \ref{obs_sed_fig} we display all the stacked data from our summer 2013 campaign. The figure shows how much the aforementioned NIR/optical/UV corrections modify the data from observed (left panel) to corrected (right panel); the emerging shape of the NIR/optical/UV SED after the corrections is remarkable and exhibits a thermal disk spectrum. In Fig. \ref{obs_sed_fig} the displayed X-ray data are only corrected for Galactic interstellar absorption and still include the effects of heavy absorption by the obscurer and the `de-ionised' warm absorber (i.e. less ionised warm absorber due to the shielding of the ionising radiation by the obscurer). Similarly, in Fig. \ref{2000_2001_2013_data} the corrected optical/UV (OM) and X-ray (EPIC-pn and RGS) data from 2000, 2001 and average summer 2013 are displayed, demonstrating that whilst in 2013 the optical/UV continuum is higher than during the unobscured epochs, the soft X-rays are heavily suppressed by the obscurer.

\subsection{Galactic interstellar reddening}
\label{reddening_sec}
To correct the optical/UV fluxes for interstellar reddening in our Galaxy, the reddening curve of \citet{Car89} was used, including the update for near-UV given by \citet{ODo94}. The {\tt ebv} model in {\tt SPEX} applies this de-reddening to the data. For \ngc the colour excess is $E(B-V) = 0.02\ \mathrm{mag}$ \citep{Sch98}. The scalar specifying the ratio of total to selective extinction $R_V \equiv A_V/E(B-V)$ was set to 3.1. In Fig. \ref{grism_fig}, the reddened continuum model is shown in dotted black line and the de-reddened one in dashed black line.

\subsection{Host galaxy stellar emission}
\label{host_gal_sect}
To take into account the contribution of starlight from the host galaxy of \ngc in our NIR/optical/UV data, we produced an appropriate model for inclusion in our spectral modelling. \citet{Ben09, Ben13} have determined the host galaxy flux at an optical wavelength for a sample of AGN (including \ngc) using \hst. However, for our observations of \ngc, this host galaxy flux was re-calculated to take into account the 5 arcsec radius circular aperture used in our processings. For the HST F550M medium-band V filter, the host galaxy flux in a 5 arcsec radius circular aperture was determined to be $6.2 \times 10^{-15}$ \ergcm with an uncertainty of $\sim 10$\% (Misty Bentz, private communication). The effective wavelength of this flux measurement is at an observed wavelength of 5580 \AA. Then, in order to calculate the host galaxy spectrum at the other wavelengths, we used a template model spectrum and normalised it to \ngc host galaxy flux at 5580 \AA. A 5 arcsec aperture takes in only the innermost few kpc of the host galaxy, and so the galaxy bulge template of \citet{Kin96} was adopted. In Fig. \ref{grism_fig}, the contribution of the host galaxy stellar model is displayed as the solid green line.

\subsection{Emission lines from BLR and NLR}
\label{BLR_NLR_sect}

To take into account emission lines produced from the BLR and NLR, we modelled the OM Optical grism and UVOT UV grism spectra, covering a range from 1895 to 6850 \AA. The simultaneity of the grism spectra with COS is important as it enables us to model the grism spectra and COS continuum points together. This wider energy band in the optical/UV helps in establishing the underlying optical/UV continuum and modelling the BLR and NLR lines. For the underlying optical/UV continuum, the Comptonisation component {\tt comt} in {\tt SPEX} (explained later in Sects. \ref{soft_excess_sect} and \ref{broadband_sect}) was used. The broad and narrow emission lines were modelled using Gaussian line profiles. For broad lines, multiple Gaussian components with different widths, but the same wavelength were added until a good fit was obtained for each line. The best fit obtained for the BLR and NLR lines was saved as a spectral model component which was included in all our broadband modelling later in Sect. \ref{broadband_sect}, to take into account the contribution of these lines in the photometric filters. The flux contribution of all the BLR/NLR lines in the grism energy range shown in Fig. \ref{grism_fig} (solid black lines) was found to be about $7.94 \times 10^{-12}$ \ergflux. 

\subsection{\ion{Fe}{ii} blended emission feature}
\label{FeII_sect}

The \ion{Fe}{ii} in the BLR produces several thousands of transitions, which result in a blended and complex spectrum between $\sim${2000--4000 \AA\ } (see e.g. \citealt{Netz83}; \citealt{Wills85}). In order to take into account the \ion{Fe}{ii} contribution in our optical data we took the model calculated by \citet{Wills85}, which is convolved to an intrinsic width of 2500 \kms. This was then imported into {\tt SPEX} as a spectral model component and convolved with the resolution of the grisms. The normalisation scaling factor of the component was left as a free parameter to allow for fitting the total \ion{Fe}{ii} flux in \ngc. The flux contribution by \ion{Fe}{ii} in \ngc was found to be about $2.79 \times 10^{-12}$ \ergflux. This component is shown in Fig. \ref{grism_fig} as part of the magenta coloured feature.

\subsection{Balmer continuum feature}
\label{Balmer_sect}

Apart from \ion{Fe}{ii} there is another significant emission component in the $\sim${2000--4000 \AA} region of AGN spectra: the Balmer continuum (see e.g. \citealt{Grandi82}). The Balmer continuum for an optically-thin emission with constant electron temperature is given by ${F_\nu ^{{\rm{BC}}} = F_\nu ^{{\rm{BE}}}{e^{ - h(\nu  - {\nu _{{\rm{BE}}}})/k{T_{\rm{e}}}}}}$  \citep{Grandi82}, where $F_\nu ^{{\rm{BC}}}$ is the Balmer continuum flux at frequency $\nu$, $F_\nu ^{{\rm{BE}}}$ is the flux at the Balmer edge $\nu_{\rm BE}$, $h$ the Planck constant, $k$ the Boltzmann constant and $T_{\rm{e}}$ the electron temperature. The Balmer edge is at theoretical wavelength of 3646 \AA. We used this model to take into account the contribution of the Balmer continuum in our optical data. The model was convolved with an intrinsic Doppler velocity broadening $\sigma_v$ and the resolution of the grisms in {\tt SPEX}. We fitted this model to the grism data for various combinations of $T_{\rm{e}}$ and $\sigma_v$ values and found the best-fit Balmer continuum component has $T_{\rm{e}}  \sim 8000$~K and $\sigma_v \sim 10000\ \kms$. We note that the purpose of fitting the Balmer continuum in this work is to obtain its total flux contribution to the optical data in order to establish the underlying continuum component, so we do not delve into the $T_{\rm{e}}$ and $\sigma_v$ parameters here which carry large uncertainties. The flux contribution by the Balmer continuum was found to be about $5.75 \times 10^{-12}$ \ergflux.
\\

\begin{figure*}[!]
\centering
\resizebox{0.85\hsize}{!}{\includegraphics[angle=0]{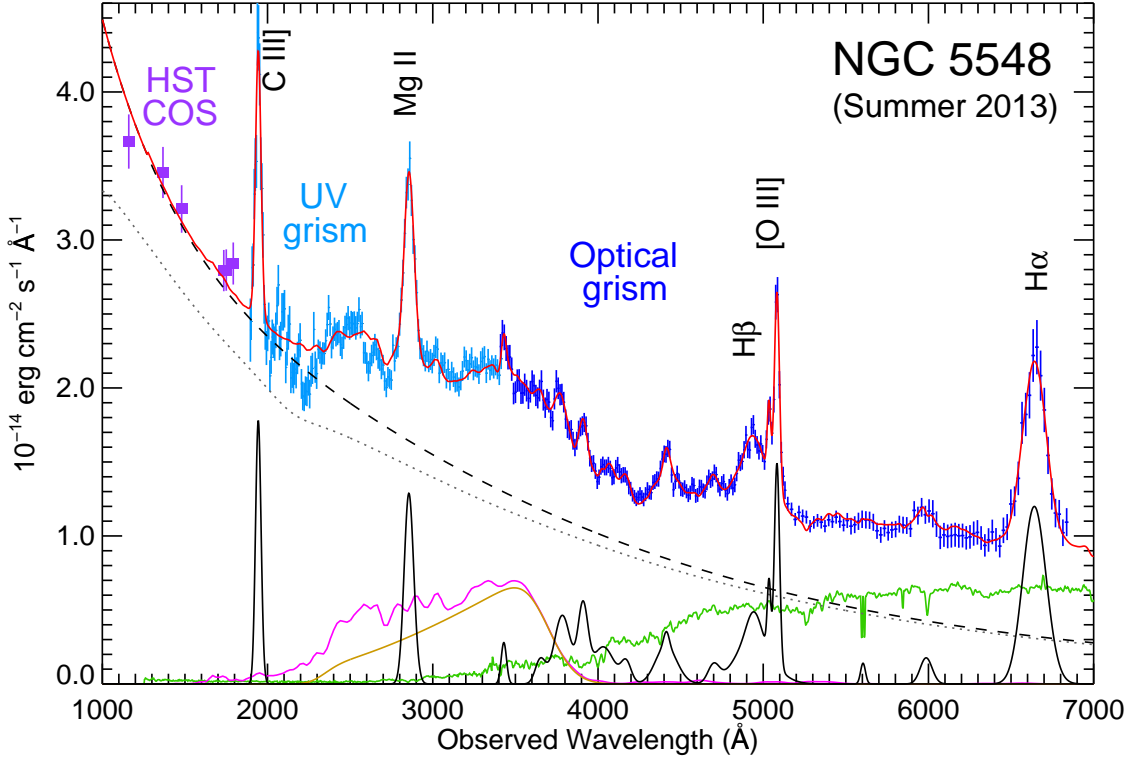}}
\vspace{-0.2cm}
\caption{De-reddened optical/UV spectrum of \ngc. The displayed data are from simultaneous UVOT UV grism spectra (shown in light blue) at 1896--3414 \AA, OM Optical grism spectra (shown in dark blue) at 3410--6836 \AA\ and the six HST COS continuum points (shown as purple squares) between 1160--1793 \AA. The displayed data are the average of the contemporaneous OM, UVOT and HST COS observations taken in summer 2013 and are corrected for Galactic reddening using the model described in Sect. \ref{reddening_sec}. The best-fit model (described in Sect. \ref{data_correct_sect}) is shown in red and various components contributing to the model are also displayed. The underlying continuum model ({\tt comt}) is the dashed black curve. The dotted black curve is the reddened version of the continuum to illustrate the correction for reddening. The broad and narrow emission line components (Sect. \ref{BLR_NLR_sect}) are shown at the bottom and some of the prominent emission lines are labelled. The broad feature in magenta is the `small-blue-bump': blended \ion{Fe}{ii} emission (Sect. \ref{FeII_sect}) with Balmer continuum (Sect. \ref{Balmer_sect}). The model for the Balmer continuum alone is shown in dark yellow below the small-blue-bump. The contribution from the host galaxy of \ngc (Sect. \ref{host_gal_sect}) is shown in green.}
\label{grism_fig}
\end{figure*}

The Balmer continuum, together with \ion{Fe}{ii} emission, form a broad and blended feature in AGN spectra dubbed the `small-blue-bump'. This is displayed in Fig. \ref{grism_fig}, in which our best-fit model to the grism spectra and HST COS continua points are shown. The small-blue-bump is the feature in magenta colour between $\sim${2000--4000 \AA}; beneath that the Balmer continuum model (without \ion{Fe}{ii} emission) is displayed in dark yellow.

%

%

\subsection{Galactic interstellar X-ray absorption}
\label{Gal_abs_sect}

The effects of the Galactic neutral absorption in the interstellar medium are included in our modelling by applying the {\tt hot} model in {\tt SPEX}. This model calculates the transmission of gas in collisional ionisation equilibrium. For a given temperature and set of abundances, the model calculates the ionisation balance and then determines all the ionic column densities by scaling to the prescribed total hydrogen column density \NH. The transmission includes both continuum and line opacity. To mimic the transmission of a neutral plasma in collisional ionisation equilibrium (such as the interstellar medium of our Galaxy), the temperature of the plasma is set to 0.5~eV. In our modelling the Galactic \ion{H}{i} column density in the line of sight to \ngc was fixed to $N_{\mathrm{H}}={1.45\times 10^{20}\ \mathrm{cm}^{-2}}$ \citep{Wakk11} with \citet{Lod09} abundances.

%
\begin{figure}[!]
\centering
\resizebox{1.025\hsize}{!}{\hspace{-0.7cm}\includegraphics[angle=0]{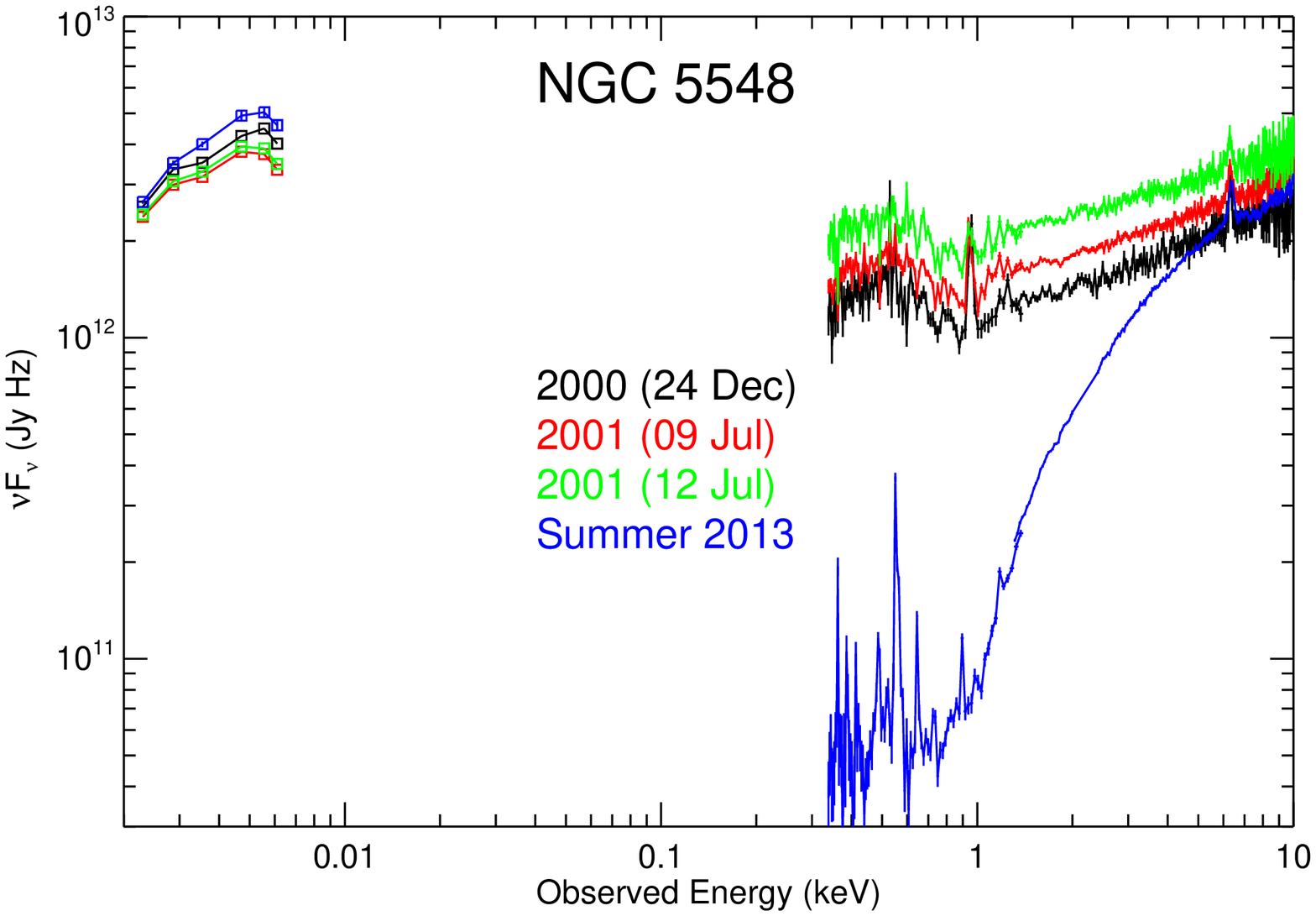}}
\caption{Optical/UV (OM) and X-ray (RGS and EPIC-pn) data of \ngc from unobscured (2000, 2001) and obscured (2013) epochs. The optical/UV data have been corrected for the effects described in Sects. \ref{reddening_sec}--\ref{Balmer_sect}. The X-ray spectra have been corrected for Galactic absorption (Sect. \ref{Gal_abs_sect}), but not for absorption by the obscurer and the warm absorber. The displayed X-ray data between 0.3--1.4 are RGS and at higher energies EPIC-pn. The X-ray spectra have been binned for clarity of presentation. In the 2013 data, the soft X-ray spectrum is the lowest one and the optical/UV data the highest one.}
\label{2000_2001_2013_data}
\end{figure}

\subsection{Absorption by the obscurer}
\label{Obsc_abs_sect}

To take into account the newly-discovered X-ray obscuration in \ngc we adopted the same model found by \citet{Kaas14}. In this model the obscurer consists of two ionisation phases, each modelled with an {\tt xabs} component in {\tt SPEX}. The {\tt xabs} model calculates the transmission through a slab of photoionised gas where all ionic column densities are linked in a physically consistent fashion through the {\tt CLOUDY} photoionisation model. The fitted parameters of an {\tt xabs} model component are the ionisation parameter ($\xi$), the equivalent hydrogen column density ($\NH$), the covering fraction $C_f$ of the absorber, its flow $v$ and turbulent $\sigma_v$ velocities. The ionisation parameter $\xi$ \citep{Tar69} is defined as 
\begin{equation}
\label{xi_eq}
\xi \equiv \frac{L}{{n_{\rm{H}}\,r^2 }}
\end{equation}
where $L$ is the luminosity of the ionising source over the 1--1000 Ryd (13.6 eV to 13.6 keV) band in $\rm{erg}\ \rm{s}^{-1}$, $n_{\rm{H}}$ the hydrogen density in $\rm{cm}^{-3}$ and $r$ the distance between the ionised gas and the ionising source in cm. The first component of the obscurer covers about 86\% of the central X-ray emitting region, with $\log \xi = -1.2$ and ${\NH = 1.2 \times 10^{22}\ \rm{cm}^{-2}}$. The second component of the obscurer covers 30\% of the X-ray source and is almost neutral ($\log \xi = -4.0$) with ${\NH = 9.6 \times 10^{22}\ \rm{cm}^{-2}}$. The parameters of the obscurer were fixed to those of \citet{Kaas14} in our broadband spectral modelling of the stacked summer 2013 data. Of course for the unobscured epochs (2000 and 2001 \xmm data), the obscurer components were excluded from our model by setting $C_f$ of both obscurer components to zero.

\subsection{Absorption by the warm absorber outflows}
\label{WA_abs_sect}

Since the obscurer is located between the central ionising source and the warm absorber, it prevents some of the ionising radiation from reaching the warm absorber. Thus the different phases of the warm absorber become less ionised, resulting in more X-ray absorption than when \ngc was unobscured. In our modelling of the obscured data we used the de-ionised warm absorber model obtained by \citet{Kaas14}, which consists of six different phases of photoionisation. Each phase is represented by an {\tt xabs} component in {\tt SPEX} with \NH, $\xi$, flow $v$ and turbulent $\sigma_v$ velocities kept frozen throughout our modelling to the values given in \citet{Kaas14}. The warm absorber phases have \NH ranging from 2 to 57 $\times 10^{20}\ \rm{cm}^{-2}$, $\log \xi$ from 0.33 to 2.67, $\sigma_v$ from 20 to 210 \kms and outflow velocities ranging from about 250 to 1220 \kms. For the 2000 and 2001 \xmm data, when the warm absorber would have been exposed to normal ionising flux, parameters of the warm absorber were taken from a re-analysis of their RGS data, using the same method as in \citet{Kaas14}. This work will be reported in a forthcoming paper on our campaign \citep{Ebre14}, where the long-term variability of the warm absorber is going to be presented.

\subsection{Soft X-ray emission lines}
\label{xray_emission_line_sect}

The 2013 RGS data clearly show the presence of several narrow emission lines and radiative recombination continua from photoionised gas \citep{Kaas14}. These emission features are more apparent in 2013 as the soft X-ray continuum is suppressed due to obscuration, but they are also present at earlier epochs. In order to take into account their flux contribution in our continuum modelling we used the same model and parameters reported in \citet{Kaas14}. The contribution of these lines to the total 0.3--2.0 keV flux is 1.4\% (in 2000), 0.8--1.3\% (in 2001) and 7.7\% (in 2013). So whilst in the unobscured epochs their contribution is rather small, in obscured epochs they contribute more to the soft X-ray flux. In another forthcoming paper on our campaign by \citet{Whew14}, a detailed study of the soft X-ray emission features as detected by RGS in \ngc is reported.

\section{The soft X-ray excess in \ngc}
\label{soft_excess_sect}

\begin{figure}[!]
\centering
\vspace{-0.6cm}
\resizebox{1.09\hsize}{!}{\hspace{-0.7cm}\includegraphics[angle=0]{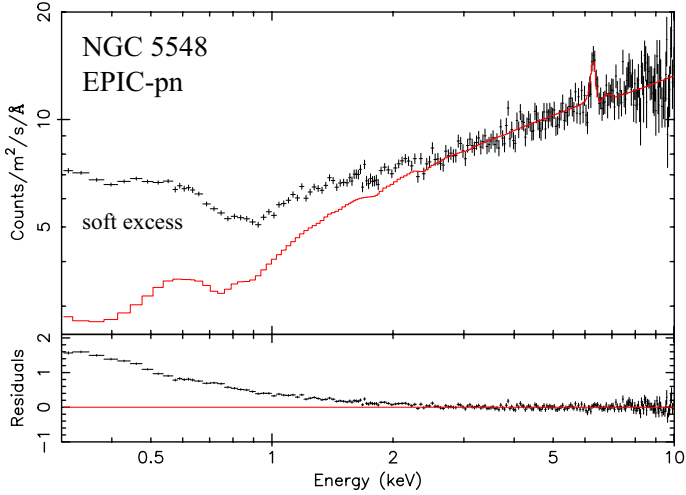}}
\vspace{-0.7cm}
\caption{EPIC-pn spectrum of \ngc from 2000, fitted above 3.0 keV with a Galactic-absorbed power-law (${\Gamma \sim 1.8}$), including a reflection component for the \FeKa. The fit is extrapolated to lower-energies displaying the presence of a soft excess below about 2 keV. The fit residuals, (observed$-$model)/model, are displayed in the bottom panel.}
\label{2000_excess_fig}
\end{figure}

The AGN `soft X-ray excess' is an excess continuum emission above the intrinsic X-ray power-law at the soft X-ray energies (below $\sim 2$~keV). Since its discovery by \cite{Sin85} in HEAO-1 observations of \mrk and also by \citet{Arn85} in \exosat observations of \object{Mrk 841}, it has been observed in many Seyfert-1 AGN. For example, in the Catalogue of AGN In the XMM-Newton Archive (CAIXA, \citealt{Bian09}), this component was commonly found in about 80\% of the AGN. In \ngc the soft excess has been previously detected (e.g. \citealt{Kaa89,Mag98}) and its presence in the unobscured state is apparent. Figure \ref{2000_excess_fig} shows the EPIC-pn spectrum obtained in 2000. The continuum above $3.0$ keV has been fitted with a Galactic-absorbed power-law component (${\Gamma \sim 1.8}$), including a reflection component for modelling the \FeKa line. The best-fit is then extrapolated to lower energies, displaying the presence of the soft excess below about 2~keV. However, the soft excess in \ngc only stands out in data before the X-ray obscuration began (i.e. $<$ 2011). Because of the newly discovered obscurer \citep{Kaas14}, the soft X-ray flux is heavily absorbed in 2013. This demonstrates that the soft excess emission is produced in a small region interior to the obscurer (e.g. the BLR) and is not from a large-scale scatterer. Yet, the soft X-ray band shows significant variability which can be attributed to both the obscurer and/or soft excess component. Therefore, for any variability study of the obscurer, the variability of the soft excess component also needs to be established. This means an appropriate model for the soft excess is required which we describe below.
\\

%

\begin{figure}[!]
\centering
\resizebox{1.02\hsize}{!}{\hspace{-0.7cm}\includegraphics[angle=0]{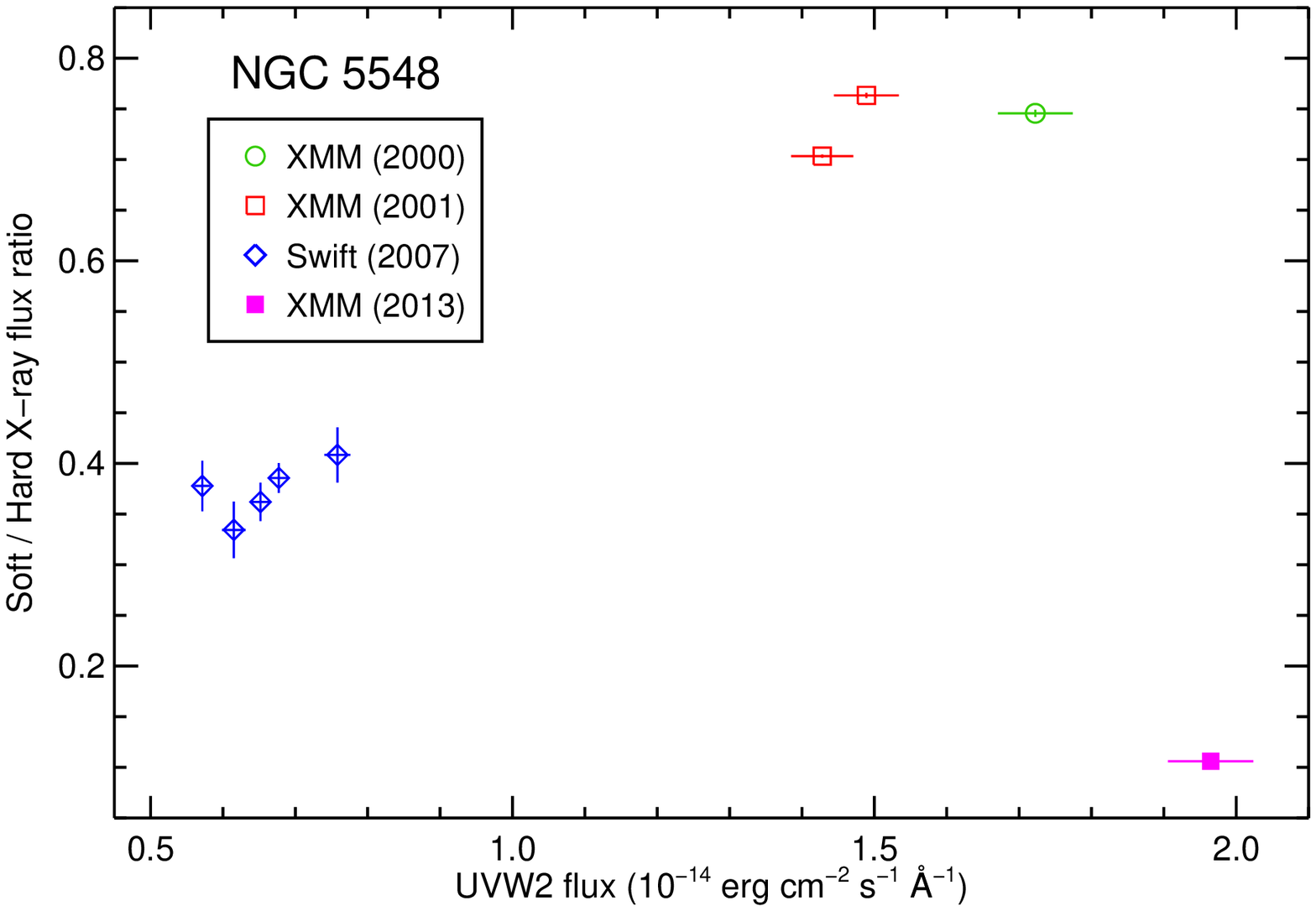}}\vspace{-0.3cm}
\resizebox{1.02\hsize}{!}{\hspace{-0.7cm}\includegraphics[angle=0]{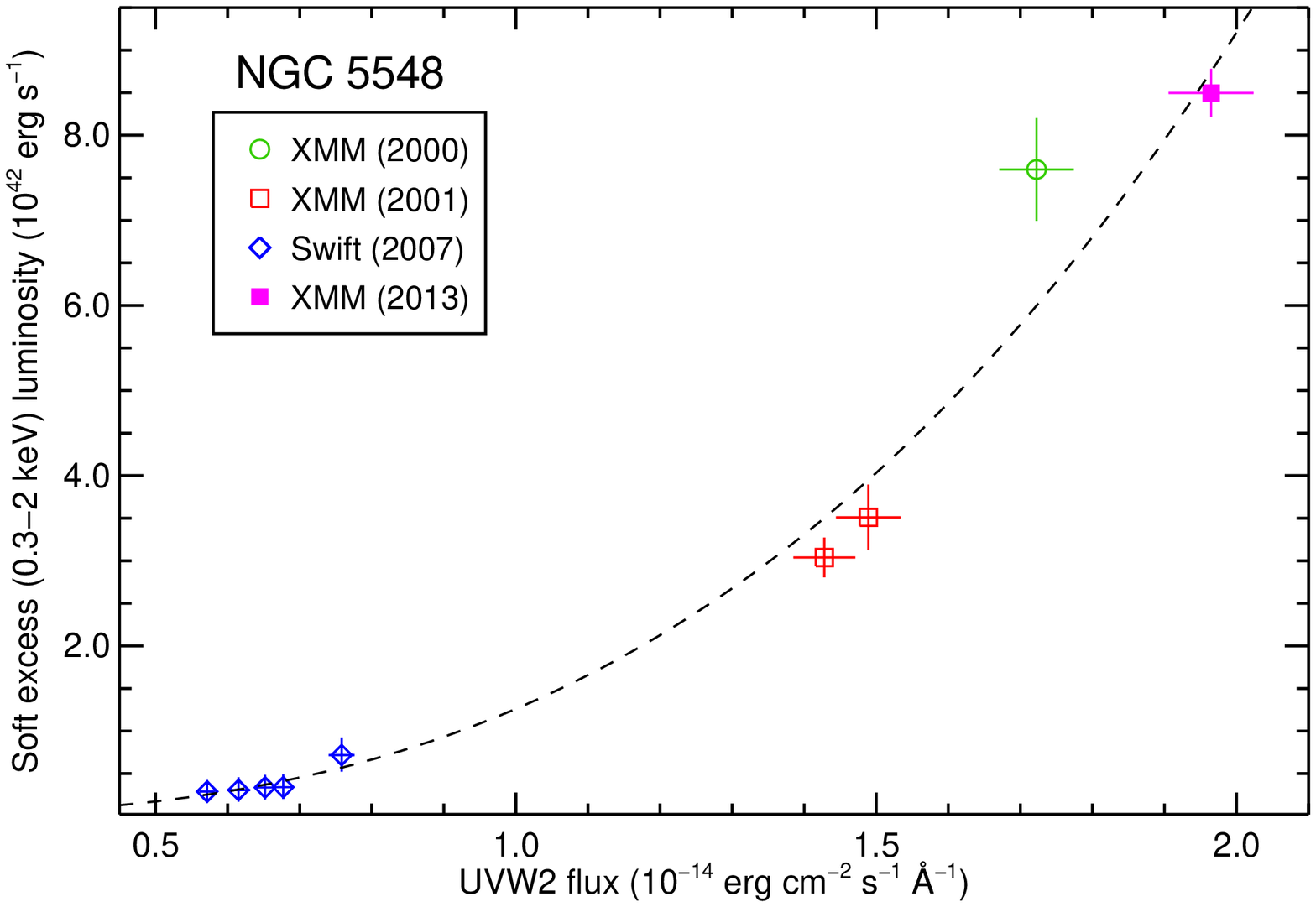}}
\vspace{-0.5cm}
\caption{{\it Top panel}: Ratio of observed soft (0.3--2.0 keV) to hard (2.0--10 keV) X-ray flux plotted versus the observed UVW2 ($2030\ \AA$) flux. {\it Bottom panel}: Intrinsic luminosity of the soft excess component between 0.3--2.0 keV plotted versus the observed UVW2 ($2030\ \AA$) flux, and fitted with the power-law function given in Eq. \ref{soft_excess_eq}.}
\label{soft_uv_rel}
\end{figure}

There are different interpretations and models in the literature for the nature of the enigmatic soft excess emission in AGN (see Sect. \ref{discuss_soft_excess}). For the case of \ngc, evidence (described below) suggests that the soft X-ray excess is primarily part of an optical/UV/soft X-ray continuum component, produced by `warm Comptonisation': up-scattering of the seed disk photons in a Comptonising corona, which is distinct from the one responsible for the hard X-ray power-law (i.e. the hot corona), and is characterised by a lower temperature and higher optical depth (e.g. \citealt{Mag98}). Firstly, without any modelling, the unobscured \ngc data (Fig. \ref{soft_uv_rel}, top panel) show that as the UV flux increases, the ratio of soft to hard X-ray flux (`softness' ratio) also increases. This indicates the UV and soft excess are related to each other. In Fig. \ref{soft_uv_rel} (top panel) the X-ray softness ratio is plotted versus the UV flux for the unobscured data (\xmm 2000 and 2001 and \swift 2007), and also for the obscured \xmm data from the average of the summer 2013 observations. The softness ratio corresponds to the observed flux, without any modelling and corrections for absorption. Remarkably, for the obscured epoch in 2013, the softness ratio is at its lowest level even though the UV flux is at its highest. The reason for this sharp drop in the softness ratio is the presence of the heavy soft X-ray absorption by the obscurer in 2013 (see Fig. \ref{2000_2001_2013_data}).
\\

%

The link between the UV and soft excess flux was also seen in \mrk \citep{Meh11}. From broadband modelling of its continuum, \citet{Meh11} and \citet{Petr13} argued warm Comptonisation to be the most reasonable explanation for the soft excess. Furthermore, warm Comptonisation has already been suggested for the soft excess in \ngc, in one of the first papers investigating this interpretation, by \citet{Mag98}. From broad-band spectral analysis of \ngc, using data from IUE, Ginga, ROSAT, CGRO and OSSE, \citet{Mag98} found that the soft excess requires a separate continuum component which can be fitted by Comptonisation of thermal photons from a cold disk ($T_{\rm seed} \sim 3.2$ eV), in a warm ($\sim 0.3$ keV), optically thick ($\tau \sim 30$) corona. The authors also found the optical/UV and soft X-ray fluxes in \ngc to be closely correlated. Therefore considering the above evidence we adopt warm Comptonisation as the appropriate model for the soft excess.
\\

In Fig. \ref{soft_uv_rel} (bottom panel) we show the soft excess luminosity in the 0.3--2.0 keV band versus the UV flux, showing the relation between the soft excess and UV. The soft excess luminosity is calculated using the model described in Sect. \ref{broadband_sect} (i.e. corrected for obscurer, warm absorber and Galactic absorption). The data in Fig. \ref{soft_uv_rel} (bottom panel) have been fitted with a function which is found to be a power-law, given by
\begin{equation}
\label{soft_excess_eq}
{L_{{\rm{soft\ excess}}}} = 1.261\, {({F_{{\rm{UVW2}}}})^{2.869}}
\end{equation}
where $L_{\rm soft\ excess}$ is the intrinsic luminosity of the soft excess component in the 0.3--2.0 keV band in units of $10^{42}\ {\rm erg}\ {\rm s}^{-1}$ and $F_{\rm UVW2}$ is the UVW2 filter flux in units of \ergcm. This flux in the UVW2 filter is at the wavelength of $2030\ \AA$ for both UVOT and OM and is the observed flux without any modifications such as de-reddening, but takes into account the optical/UV spectrum shape of \ngc (shown in Fig. \ref{grism_fig}) in the UVW2 filter bandpass. We further discuss the soft excess in Sect. \ref{discuss_soft_excess}.

\section{Broadband continuum modelling of \ngc}
\label{broadband_sect}

Here we describe our continuum modelling of the \ngc data. The archival \xmm data from 2000 and 2001 and the stacked 2013 data from our campaign have been used. The archival data are from epochs before the source became obscured and its spectrum more complex, and so are useful in establishing the underlying continuum of \ngc. The stacked 2013 data refer to stacked \xmm data (Obs. 1--12) and the corresponding simultaneous \nustar, \integral, HST COS, \wise (I and R filters) observations collected during the summer 2013 campaign. The stacked observations are all from a period when the X-ray flux was at a stable low level with only moderate variability between observations, such that it does not have an adverse effect on the results obtained from the stacked spectrum. The stacking of spectra was performed using the {\tt mathpha}, {\tt addrmf}, and {\tt addarf} tools of the {\tt HEASOFT} package, except for RGS where the stacking was done using the procedure described in Appendix \ref{rgs_sect}. All the X-ray spectra used in our fitting have been optimally binned in {\tt SPEX} (the {\tt obin} command) taking into account the instrumental resolution and count statistics of the source as described in the {\tt SPEX} manual. In our broadband spectral modelling we included all the model components discussed in Sect. \ref{data_correct_sect}, which take into account: Galactic reddening, host galaxy stellar emission, BLR and NLR emission lines, blended \ion{Fe}{ii} and Balmer continuum, Galactic X-ray absorption, the warm absorber, soft X-ray emission lines and finally absorption by the obscurer (for the 2013 data). 
\\

As explained in Sect. \ref{soft_excess_sect}, warm Comptonisation is a reasonable and likely model for the soft excess in \ngc. So in our broadband modelling of the continuum, we applied the {\tt comt} model in {\tt SPEX} to simultaneously fit the NIR/optical/UV and soft X-ray data. The {\tt comt} model is based on the Comptonisation model of \citet{Tit94}, but with improved approximations for the parameter $\beta (\tau)$ which characterises the photon distribution over the number of scatterings which the soft photons undergo before escaping the plasma (for more details on this see the {\tt SPEX} manual). The up-scattering Comptonising plasma was set to have a disk geometry and its parameters are: the temperature of the seed photons $T_{\rm seed}$, the electron temperature of the plasma $T_{\rm e}$, its optical depth $\tau$, and normalisation of the component. We note the parameterisation of the warm corona is model-dependent and different Comptonisation models can lead to slightly different $T$ and $\tau$ values. All models however give qualitatively the same results, i.e. low-$T$ and high-$\tau$ values and reproduce well the characteristic spectral shape of the soft excess.
\\

\begin{figure*}[!]
\centering
\resizebox{0.92\hsize}{!}{\includegraphics[angle=0]{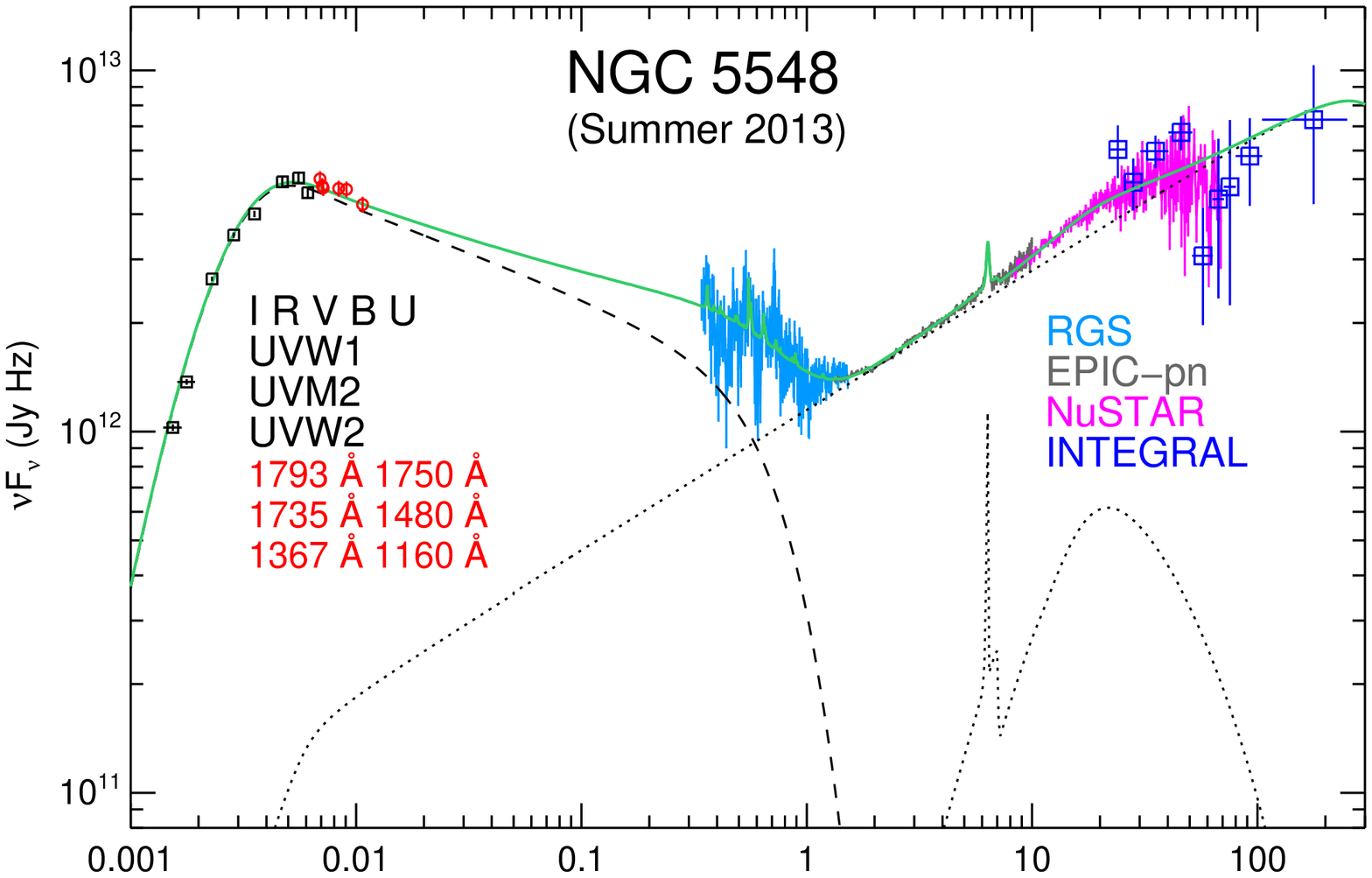}}\vspace{-1.3 cm}
\resizebox{0.92\hsize}{!}{\includegraphics[angle=0]{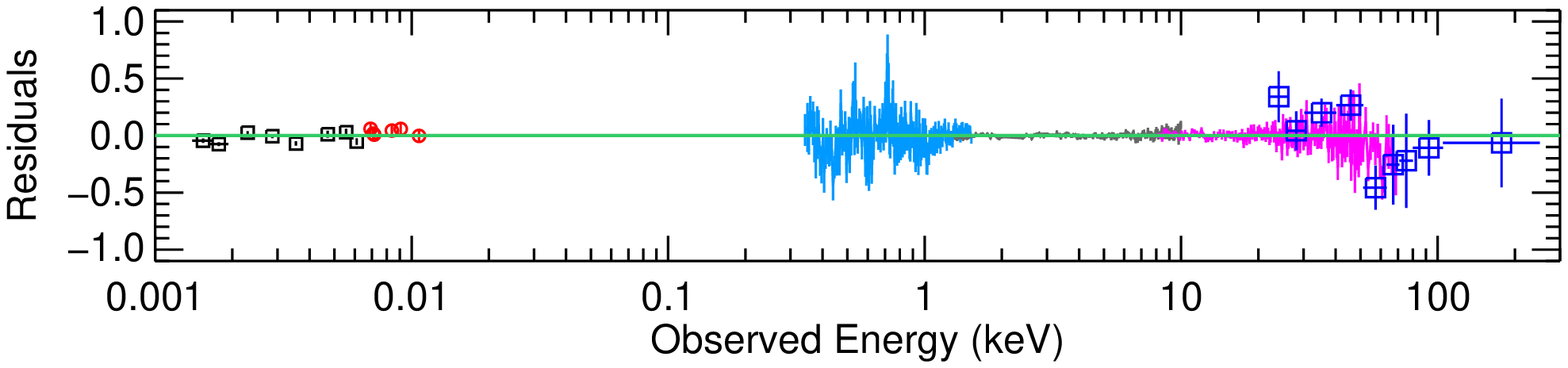}}
\caption{Best-fit broadband continuum model for the stacked summer 2013 data of \ngc as described in Sect. \ref{broadband_sect}. All the displayed data, labelled in the figure, have been corrected for the emission and absorption effects described in Sect. \ref{data_correct_sect}, including obscuration and warm absorption, required for determining the underlying continuum. The soft X-ray emission features seen in the RGS are shown on the continuum model for display purposes. The {\tt comt} warm Comptonisation component (modelling the soft excess) is shown in black dashed-line, with the power-law ({\tt pow}) and reflection ({\tt refl}) components in black dotted-line form. The broadband model spectrum is represented in a solid green line. Residuals of the fit, defined as (data $-$ model) / model, are displayed in the bottom panel.}
\label{2013_SED_continuum}
\end{figure*}

In addition to the {\tt comt} component which models the NIR/optical/UV continuum and the soft X-ray excess, a power-law component ({{\tt pow} in {\tt SPEX}}) was also included in our model. This component models the primary hard X-ray continuum which has a power-law spectral shape and is thought to be produced by inverse Compton scattering in an optically thin (${\tau \sim 1}$), hot (${T_{\rm{e}} \sim 100\ {\rm keV}})$ corona (e.g. \citealt{Suny80,Haar93}). The cut-off power-law is known to provide a reasonable approximation for the Comptonisation spectrum (e.g. \citealt{Petr00}). The high-energy exponential cut-off of the power-law was set to 400~keV in our modelling; for details of the high-energy cut-off parameter search in \ngc see \citet{Ursin14}. We note the seeming decline in about 50--80 keV in Fig. \ref{2013_SED_continuum} is background dominated and not statistically significant. The actual high-energy cut-off is at higher energies ($\sim$300--400~keV). At low energies, the power-law was smoothly broken before overshooting the energies of the seed photons from the disk. The break energy was fixed to the peak energy of the {\tt comt} component. The precise energy, which is rather uncertain, was found to have negligible effect in the NIR/optical/UV modelling of the \ngc data, as the {\tt comt} component is overly dominant at these energies (see Fig. \ref{2013_SED_continuum}).
\\

To model the \FeKa line and the reflection spectrum at hard X-ray energies we used the reflection model {\tt refl} in {\tt SPEX}. This model, provided by Piotr Zycki, computes the reflected continuum and the corresponding \FeKa line from a constant density X-ray illuminated atmosphere. It incorporates the Compton-reflected continuum of \citet{Magd95} and the \FeKa line model of \citet{Zyck94}. The parameters of the model were fixed to those reported in \citet{Kaas14} with no relativistic broadening. We fitted the normalisation and photon index $\Gamma$ of its illuminating power-law in our modelling to determine the cold reflection from distant material. However, for the 2000 and 2001 observations (in which no hard X-ray \nustar and \integral data were available), $\Gamma$ of {\tt refl} was frozen to the value found from 2013 data ($\Gamma = 2.2 \pm 0.1$) and only the normalisation was fitted in order to make a good fit to the \FeKa. The power-law normalisations of the reflection component were found to be $4.4 \pm 0.4$ (for 2000), $3.5 \pm 0.3$ (for 9 July 2001), $4.4\pm0.4$ (for 12 July 2001) and $3.0 \pm 0.4$ (for stacked 2013) in units of $10^{52}$ photons $\mathrm{s}^{-1}$ $\mathrm{keV}^{-1}$ at 1 keV. 
\\

In our broadband spectral modelling, $T_{\rm seed}$, $T_{\rm e}$, $\tau$ and normalisation of {\tt comt}, as well as $\Gamma$ and normalisation of {\tt pow} were fitted. At the last step in fitting the 2013 data, we tested if freeing parameters of the obscurer from their initial values \citep{Kaas14} provides any improvement to the fit. However, the obscurer parameters remained within 1-$\sigma$ errors and the goodness of the fit was not significantly improved. Therefore, for consistency we have kept the obscurer parameters fixed to the values reported by \citet{Kaas14}. Figure \ref{2013_SED_continuum} shows our broadband model fit to the stacked 2013 data. The contributions from individual continuum components ({\tt comt}, {\tt pow} and {\tt refl}) are also displayed in the figure. In Table \ref{comt_table} the best-fit parameters of our continuum model are given. The observed fluxes and intrinsic luminosities in various energy bands, calculated from our model, are given in Table \ref{lum_table}. In Fig. \ref{2000_2001_2013_continuum} we compare the intrinsic broadband continuum SED at unobscured and obscured epochs. In Fig. \ref{obscured_SED_fig} we show the broadband SED in 2013 before and after absorption by the obscurer.

%

%
\begin{table*}[!tbp]
\vspace{0.1 cm}
\caption{Best-fit parameters of the broadband continuum model for 2000, 2001 and stacked 2013 data of \ngc.}
\label{comt_table}
\small
\setlength{\extrarowheight}{3pt}
\begin{minipage}[t]{\hsize}
\renewcommand{\footnoterule}{}
\centering
\begin{tabular}{c | | l l l l | l l | c}
\hline \hline
 &  \multicolumn{4}{c|}{Warm Comptonisation ({\tt comt}):} & \multicolumn{2}{c|}{Power-law ({\tt pow}):} \\ 
Observation & $T_\mathrm{seed}$\,\footnote{\,Temperature of the seed photons in eV.} & $T_\mathrm{e}$\,\footnote{\,Warm corona temperature in keV.}  & $\tau$\,\footnote{\,Optical depth.} & Norm\,\footnote{\,$10^{55}$ photons $\mathrm{s}^{-1}$ $\mathrm{keV}^{-1}$.} & $\Gamma$\,\footnote{\,Photon index.} & Norm\,\footnote{\,$10^{51}$ photons $\mathrm{s}^{-1}$ $\mathrm{keV}^{-1}$ at 1 keV.} & C-stat / Expected C-stat \\
\hline

2000 (24 Dec) 		& $0.74 \pm 0.03$ 		& $0.15 \pm 0.01$ 	& $22.7 \pm 0.9$ 	& $6.3 \pm 0.5$ 	& $1.82 \pm 0.01$ 	& $5.94 \pm 0.08$ 	& $1544\, /\, 1353$ \\
2001 (09 July) 		& $0.71 \pm 0.03$ 		& $0.14 \pm 0.01$ 	& $22.1 \pm 0.8$ 	& $5.2 \pm 0.4$ 	& $1.81 \pm 0.01$ 	& $7.28 \pm 0.04$ 	& $1860\, /\, 1372$ \\
2001 (12 July) 		& $0.58 \pm 0.05$ 		& $0.13 \pm 0.01$ 	& $24.3 \pm 1.0$ 	& $6.4 \pm 0.7$ 	& $1.88 \pm 0.01$ 	& $10.3 \pm 0.08$ 	& $1790\, /\, 1362$ \\
2013 (Summer) 	& $0.80 \pm 0.02$ 		& $0.17 \pm 0.01$ 	& $21.1 \pm 0.5$ 	& $6.0 \pm 0.2$ 	& $1.61 \pm 0.01$ 	& $4.72 \pm 0.10$ 	& $2666\, /\, 1664$ \\

\hline
\end{tabular}
\end{minipage}
\end{table*}

%
\begin{table*}[!tbp]
\caption{Observed fluxes and intrinsic luminosities of \ngc for 2000, 2001 and stacked 2013 data. The fluxes ($F$) are given in units of ${10^{-11}\ \rm{erg\ cm}^{-2}\ \rm{s}^{-1}}$ and the intrinsic luminosities ($L$) are in $10^{43}\ {\rm erg}\ {\rm s}^{-1}$. The observed fluxes include all the absorptions described in Sects. \ref{Gal_abs_sect}, \ref{Obsc_abs_sect} and \ref{WA_abs_sect}. The host galaxy stellar emission (Sect. \ref{host_gal_sect}) is excluded from the reported flux and luminosities.}
\label{lum_table}
\small
\setlength{\extrarowheight}{3pt}
\begin{minipage}[t]{\hsize}
\renewcommand{\footnoterule}{}
\centering
\begin{tabular}{c || c c | c c | c c | c c}
\hline \hline
Observation & $F_{\rm 0.3-2\ keV}$ & $L_{\rm 0.3-2\ keV}$  & $F_{\rm 2-10\ keV}$  & $L_{\rm 2-10\ keV}$ & $F_{\rm 0.001-10\ keV}$  & $L_{\rm 0.001-10\ keV}$ & $F_{\rm 1-1000\ Ryd}$ & $L_{\rm 1-1000\ Ryd}$  \\
\hline
2000 (24 Dec) 		& $2.34$		& $2.53$		& $3.14$		& $2.17$		& $13.1  $		& $18.4$ & 	$6.50$	& 	$12.2$	 \\ 		
2001 (09 July) 		& $2.72$ 		& $2.47$		& $3.87$		& $2.67$		& $13.6  $		& $15.8$ & 	$8.04$	& 	$10.9$	 \\
2001 (12 July) 		& $3.74$ 		& $3.44$		& $4.90$		& $3.39$		& $16.5  $		& $19.6$ &	$10.5$	& 	$14.4$	 \\
2013 (Summer) 	& $0.30$ 		& $2.21$		& $2.80$		& $2.35$		& $10.5  $		& $17.9$ & 	$4.09$	& 	$11.7$	 \\

\hline
\end{tabular}
\end{minipage}
\end{table*}

\section{Thermal stability analysis of photoionised gas}
\label{stability_sect}

Having developed a broadband continuum SED model for the unobscured (2000 and 2001) and obscured (summer 2013) epochs, we investigated the effects of different SEDs and X-ray obscuration on the ionisation balance of gas. This is important for investigating the impact of the ionising SED on the obscurer and subsequently, after heavy absorption, on the warm absorber outflows.
\\

%
\begin{figure}[!]
\centering
\resizebox{1.025\hsize}{!}{\hspace{-0.7cm}\includegraphics[angle=0]{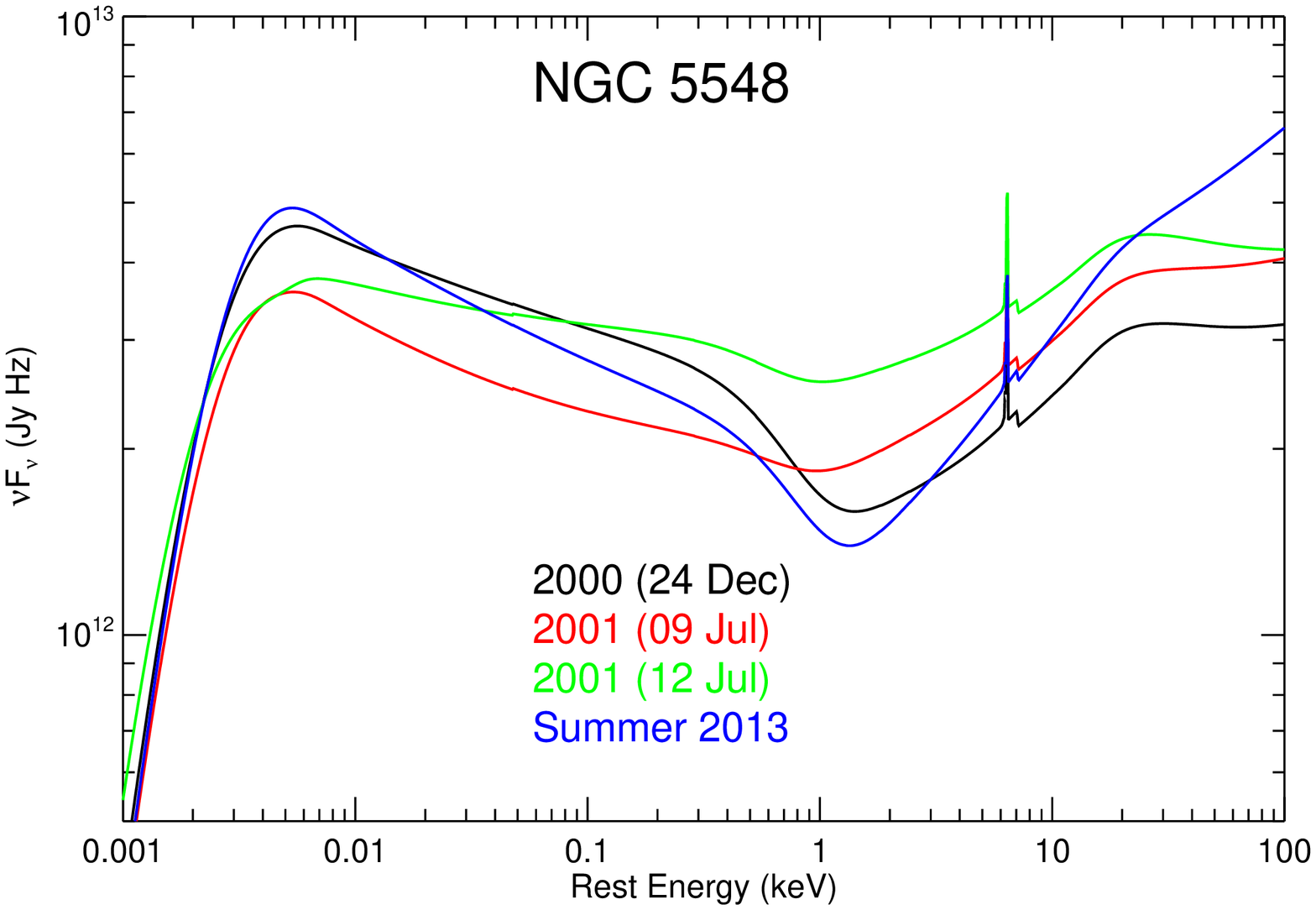}}
\vspace{-0.5cm}
\caption{The broadband continuum SED models for the unobscured (2000, 2001) and obscured (2013) epochs of \ngc, obtained in Sect. \ref{broadband_sect}.}
\label{2000_2001_2013_continuum}
%
\centering
\resizebox{1.025\hsize}{!}{\hspace{-0.7cm}\includegraphics[angle=0]{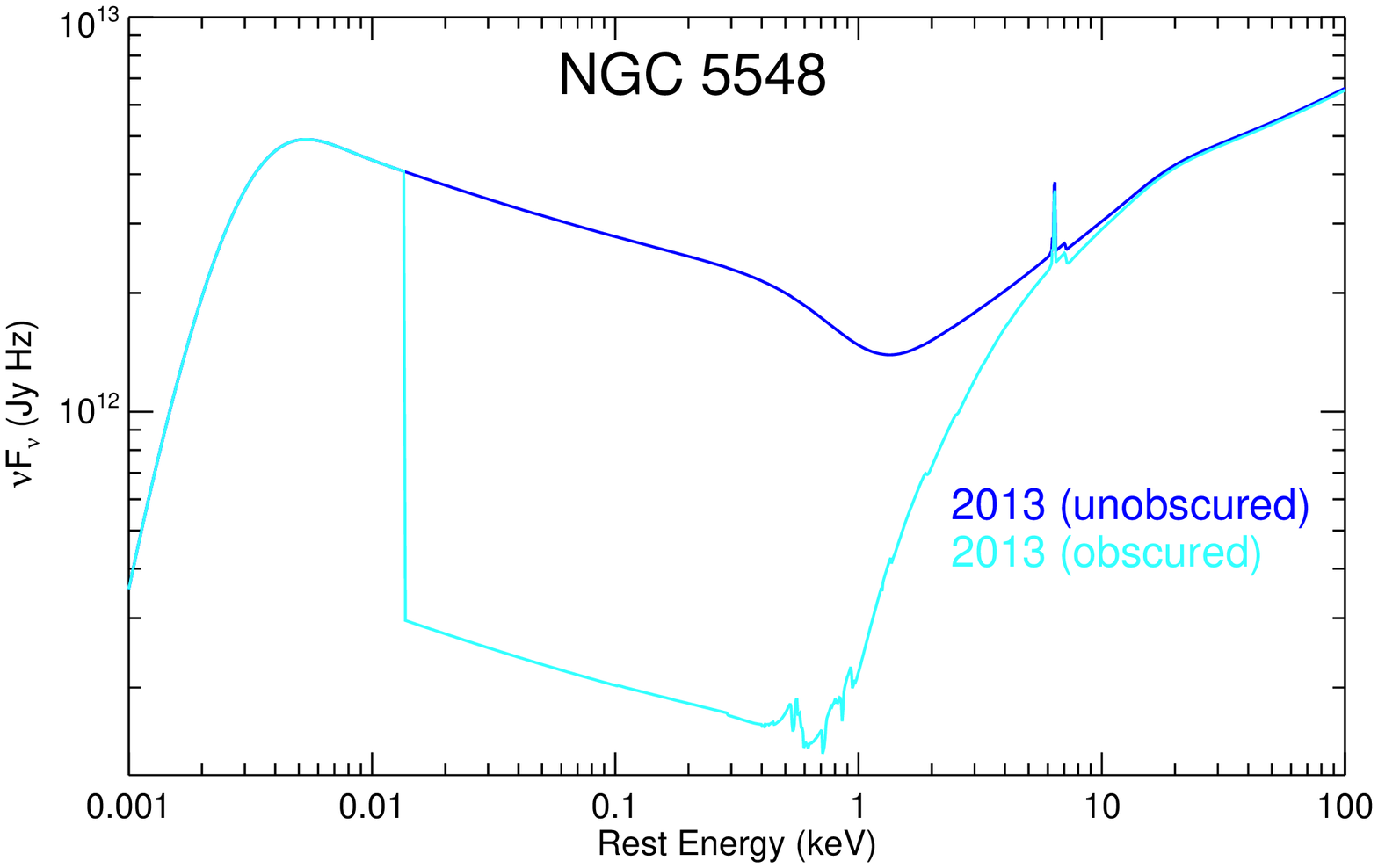}}
\vspace{-0.5cm}
\caption{The SED model for the 2013 epoch, displayed before ({\it dark blue}) and after ({\it light blue}) absorption of the continuum by the obscurer. The ionising luminosity $L_{\rm 1-1000\ Ryd}$ is $11.7 \times 10^{43}\ {\rm erg}\ {\rm s}^{-1}$ before obscuration and $3.5 \times 10^{43}\ {\rm erg}\ {\rm s}^{-1}$ after obscuration.}
\label{obscured_SED_fig}
\end{figure}

%
\begin{figure}[!]
\centering
\vspace{-0.5cm}
\resizebox{1.1\hsize}{!}{\hspace{-0.9cm}\includegraphics[angle=0]{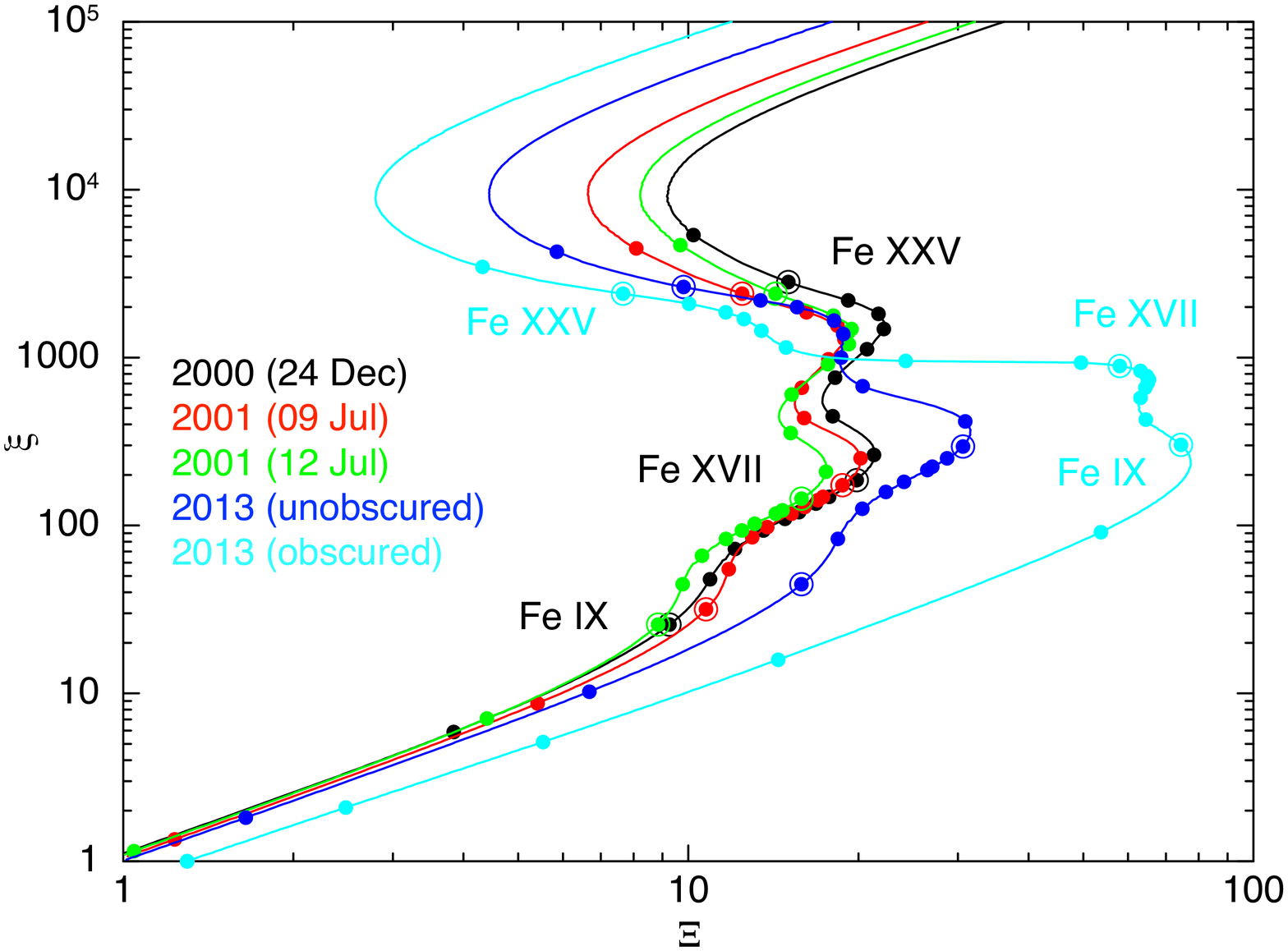}}\vspace{-1.2cm}
\resizebox{1.1\hsize}{!}{\hspace{-0.9cm}\includegraphics[angle=0]{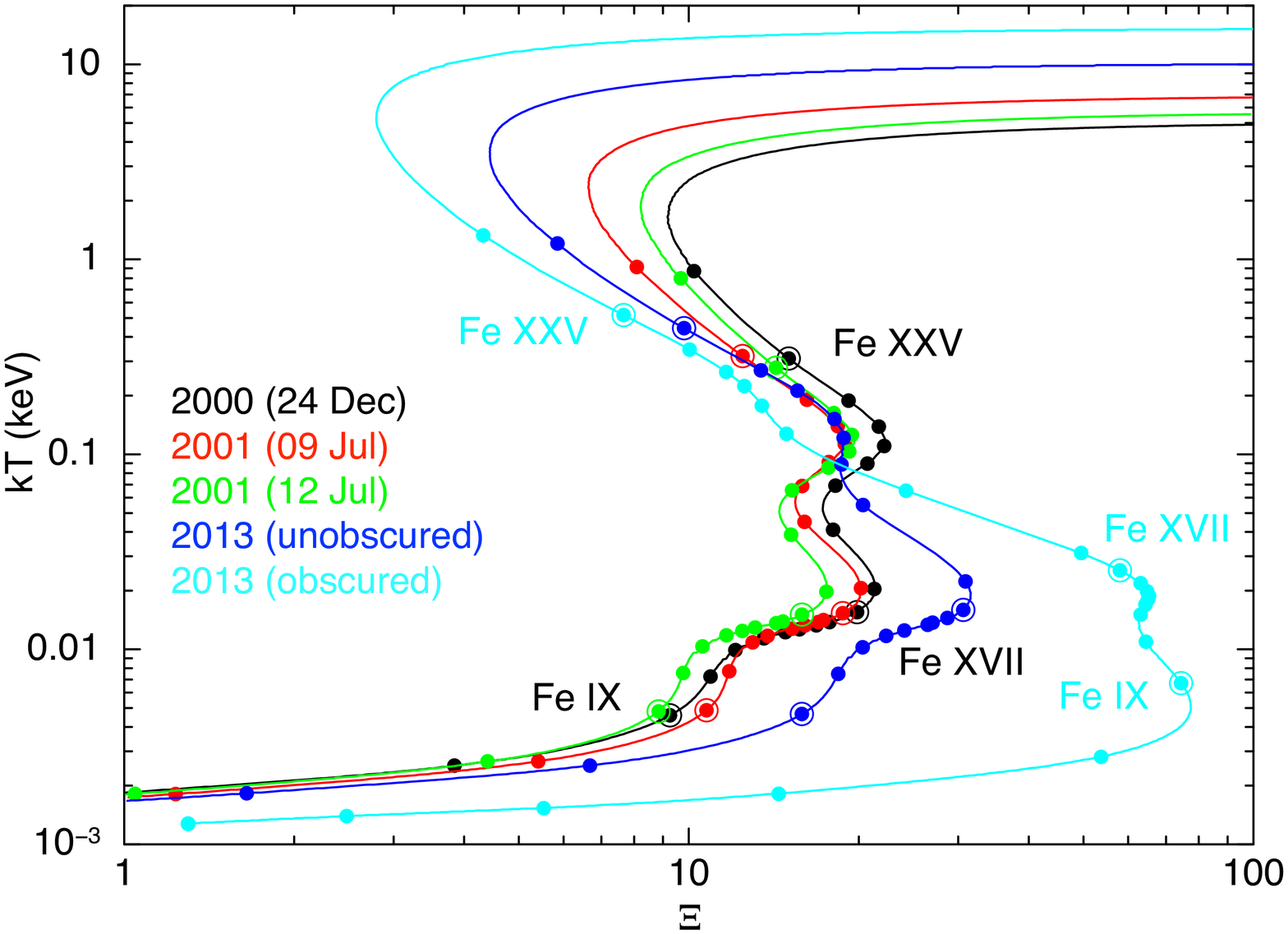}}
\vspace{-0.7cm}
\caption{Thermal stability curves for photoionised gas in \ngc, showing $\xi$ plotted versus $\Xi$ ({\it top panel}) and temperature $kT$ plotted versus $\Xi$ ({\it bottom panel}). They have been calculated for SEDs (displayed in Figs. \ref{2000_2001_2013_continuum} and \ref{obscured_SED_fig}) derived from our broadband modelling of the continuum as described in Sect. \ref{broadband_sect}. The circles on the stability curves indicate the positions at which ionic column densities of Fe peak on the ($\Xi$, $\xi$) and ($\Xi$, $T$) phase space. To guide the eye, the location of \ion{Fe}{ix}, \ion{Fe}{xvii}, \ion{Fe}{xxv} have been indicated with open circles around filled circles.}
\label{stab_curves}
\end{figure}

The ionising SED determines the ionisation balance and thermal stability of photoionised gas, such as warm absorbers in AGN. X-ray photoionised plasma can be thermally unstable in certain regions of the ionisation parameter space. This can be investigated by means of producing the thermal stability curves (also referred to as S-curves or cooling curves). These are plots of either the temperature of the plasma, or the ionisation parameter $\xi$, versus the pressure form of the ionisation parameter, $\Xi$, introduced by \citet{Kro81}. The parameter $\Xi$, which describes the ionisation equilibrium, is defined as $\Xi  \equiv {F}/{{n_{\rm{H}}\, c\, kT}}$, where $F$ is the flux of the ionising source between 1--1000 Ryd (in $\rm{erg}\ \rm{cm}^{-2}\ \rm{s}^{-1}$), $k$ the Boltzmann constant, $T$ the temperature and $n_{\rm{H}}$ is the hydrogen density in ${\rm{cm}}^{-3}$. Here, $F/c$ is the pressure of the ionising radiation. Taking $F = L / 4\pi r^{2}$ and using $\xi \equiv {L}/{{n_{\rm{H}}\,r^2 }}$, $\Xi$ can be expressed as
\begin{equation}
\label{big_xi_eq_v2}
\Xi  = \frac{L}{{4\pi r^2\, n_{\rm{H}}\, c\, kT}} = \frac{\xi }{{4\pi \, c\, kT}} \approx 19222\ \frac{\xi }{T}
\end{equation}

On the curve itself, the heating rate is equal to the cooling rate, so the gas is in thermal balance. To the left of the curve, cooling dominates over heating, whilst to the right of the curve, heating dominates over cooling. On branches of the S-curve which have positive gradient, the photoionised gas is thermally stable. This means small perturbations upward in temperature increase the cooling, whereas small perturbations downward in temperature increase the heating. However, on the branches with negative gradient, the photoionised gas is thermally unstable, in this case ${\left( {{{\partial (C - H)}}/{{\partial T}}} \right)_\Xi } < 0$, where $C$ represents the complete set of cooling processes and $H$ represents the complete set of heating processes. Therefore on the unstable branches, a small perturbation upward in temperature increases the heating relative to the cooling, causing further temperature rise, whereas a small perturbation downward increases the cooling. The thermal stability curve is highly sensitive to the SED of the ionising source and to the elemental abundances of the photoionised gas.
\\

We have computed the cooling curves (Fig. \ref{stab_curves}) corresponding to the different ionising SEDs of \ngc (Figs. \ref{2000_2001_2013_continuum} and \ref{obscured_SED_fig}). To obtain $\Xi(T)$ and $\Xi(\xi)$, we ran {\tt CLOUDY}\footnote{\url{http://www.nublado.org}} \citep{Fer98} version C13.01 for a photoionised gas with \citet{Lod09} abundances irradiated with the SEDs obtained in Sect \ref{broadband_sect}. In Fig. \ref{stab_curves} the points on the cooling curves where the Fe ions peak have been indicated. Interestingly, the highly-ionised ions of Fe fall on an unstable branch of the curves for all five cases. Moreover, the shielding by the obscurer in \ngc has a significant impact on the ionisation balance and thermal stability of photoionised gas located further out and produces a large unstable branch on the curves. We further discuss the cooling curves results in Sect. \ref{discuss_stability}.

\section{Discussion}
\label{discussion}

\subsection{The nature of the NIR/optical/UV continuum and the soft X-ray excess}
\label{discuss_soft_excess}

The NIR/optical/UV continuum of \ngc, modelled using simultaneous data from about 8060 \AA\ to 1160 \AA, can be modelled as being one single continuum component. This result has been found in both local modelling of the optical/UV grisms and HST COS data (Sect. \ref{data_correct_sect} and Fig. \ref{grism_fig}), and broadband modelling of the continuum (Sect. \ref{broadband_sect} and Fig. \ref{2013_SED_continuum}). This single continuum component is modelled by a warm Comptonisation component ({\tt comt}) which, as well as fitting the NIR/optical/UV continuum, also fits the soft X-ray excess emission. There are no additional continuum components detected in the NIR/optical/UV energies. We emphasise the importance of fully taking into account various non-continuum processes (described in Sect. \ref{data_correct_sect}) which make a significant flux contribution in broadband photometric filters. For example, emission from the blended \ion{Fe}{ii} and Balmer continuum (Sects. \ref{FeII_sect} and \ref{Balmer_sect}) contributes to 27\% of the U filter flux, and emission from the host galaxy (described Sect. \ref{host_gal_sect}) contributes to about 48\% of the V filter flux. The host galaxy contribution relative to the continuum becomes greater in the R and I photometric filters, contributing to 70\% and 79\% of the observed flux, respectively. So if only photometric filter data are fitted without taking into account the actual spectrum and secondary emission processes, one may erroneously attribute an `excess emission' to an additional continuum component. 
\\

Warm Comptonisation has been found to be a viable explanation for the soft excess in some AGN based on broadband spectral modelling or timing analyses - see e.g. \citet{Mag98,Meh11,Petr13,Jin13,Noda13,DiGesu14a,Arev14,Matt14}. The soft excess has also been attributed to reprocessed emission caused by relativistically-blurred, ionised-disk reflection in some AGN (in particular NLSy1s) based on observations of relativistically broadened Fe profiles and time lags between soft and hard X-rays - see e.g. \citet{Ros93,Ros05,Cru06,Fabi09,Fabi13,Walt13}. In this work, the correlation between UV and soft X-rays long-timescale variability and the results of our broadband continuum modelling of \ngc data suggest the origin of the soft excess is consistent with the warm Comptonisation interpretation. This was made possible by the remarkable simultaneous data from our summer 2013 campaign covering a wide energy range from NIR (8060 \AA) to hard X-rays (200 keV) and also the archival (2000, 2001 and 2007) data when the source was not obscured. Our results suggest the soft excess is primarily the tail of the Comptonised disk component extending from NIR to soft X-rays. We note that although the long-term variability of the soft excess over several years can be explained by changes in the strength of the Comptonised disk emission, on short timescales reprocessed emission from reflection can also possibly contribute to any rapid variability of the soft excess. Indeed, short-timescale lag between soft and hard X-rays has been detected in \ngc while the source was unobscured \citep{DeMarc13}, which may be interpreted as some contribution to soft excess rapid variability from reprocessing. However, the combination of the hard X-ray data of \nustar and \integral with EPIC-pn spectroscopy of the \FeKa line show the reflection spectrum and the \FeKa line energy and its narrow width are consistent with reflection from distant neutral material \citep{Ursin14, Cappi14}. From our best-fit model the contribution of the neutral {\tt refl} component to the soft excess is only $\sim$ 1\%. Although there is no spectroscopic evidence of an ionised reflection in \ngc, by invoking an additional ionised {\tt refl} component to our model with $\log \xi = 2.0$ and relativistic blurring, the contribution of the ionised reflection component to the soft excess is estimated to be about 2--4\%, with the rest of the soft excess emission being produced by the warm Comptonisation.
\\

The direct relation between the UV flux and the strength of the soft excess (Fig. \ref{soft_uv_rel}) suggests they are related and primarily part of the same emission component driving the UV variability, which is well modelled as warm Comptonisation in \ngc. A purely thermal disk blackbody spectrum however is not viable as it does not reach the soft X-rays. Furthermore, the correlated variability of the soft excess component with the UV disk emission suggests the soft excess is `real' emission and not an artefact of any relativistically-blurred absorption. Such X-ray absorbing winds are largely transparent at wavelengths longer than the Lyman limit, and thus cannot contribute to the optical/UV/X-ray correlated variability. In this work, all the X-ray absorption has been fully taken into account based on RGS spectroscopy of their features. In a recent study by \citet{Done12}, the warm Comptonisation origin for the soft excess has been tested by further modelling. The authors built an energetically self consistent accretion flow model: an optically-thick accretion disk is present down to a radius $r_{\rm{corona}}$, below which the accretion power is shared between an optically-thick warm corona and an optically-thin hot one, with both coronae Compton up-scattering the disk emission. In this proposed picture, our fit of the entire NIR/optical/UV continuum and the soft excess, without the need for an additional thermal component from the disk, suggests that the warm corona completely cover the optically-thick accretion disk, with a geometry similar to the one proposed for \mrk by \citet{Petr13}. We emphasise the importance and usefulness of simultaneous multi-wavelength data (like those obtained in \mrk and \ngc campaigns) in investigating the nature of the soft excess, as well as other intrinsic components of AGN spectra. 

\subsection{The continuum at unobscured and obscured epochs}
\label{discuss_obsc}

The basic nature of the underlying broadband continuum in both unobscured and obscured epochs is unchanged. However, the value of $\Gamma$ of the power-law in the stacked 2013 spectrum is about 1.61, which is lower than those from archival 2000 and 2001 ranging over 1.81--1.88. We note this spectral hardening observed in 2013 is not an artefact caused by the spectral fitting of the complex absorption by the obscurer. The hard X-ray data obtained by \nustar and \integral confirm this spectral hardening \citep{Ursin14}. Additionally, \swift data from unobscured epochs of 2005 and 2007 show that $\Gamma$ has previously been between 1.5--1.6 \citep{Meh14b}. The long-term variability of $\Gamma$, presented in \citet{Meh14b}, shows that even at some obscured epochs (2012 and 2014) $\Gamma$ was consistently higher, ranging typically between 1.7--1.9, whereas during the time of our summer 2013 \xmm campaign it was lower at 1.5--1.6. So the low $\Gamma$ of 1.61 found for the stacked 2013 spectrum is not related to the fact that the source is obscured and that the full effect of its absorption was not correctly taken into account.

\subsection{Effect of obscuration on ionisation and thermal stability of photoionised gas}
\label{discuss_stability}

Determining the shape of the ionising SED using a proper broadband continuum model is important in assessing the ionisation balance of photoionised gas. For instance, \citet{Chak12} have shown that just the temperature of the accretion disk and strength of the soft X-ray excess component can significantly influence the structure and stability of the warm absorbers in AGN. In Sect. \ref{stability_sect} we presented the cooling curves of photoionised gas in \ngc based on broadband ionising SEDs of the source for unobscured and obscured cases. On the cooling curves, the heating rate is equal to the cooling rate so the gas is in thermal equilibrium. This is reached when photoionisation and Compton heating are balanced by recombination and Compton cooling, amongst other less dominant heating/cooling mechanisms, which are all taken into account by the {\tt CLOUDY} code. As shown in Fig \ref{stab_curves}, the cooling curves from unobscured 2000 and 2001 SEDs display similar shapes. But the 2013 curves and in particular the obscured version deviate strongly from those corresponding to the unobscured SEDs.
\\

All five cooling curves in Fig \ref{stab_curves} show that up to $\xi$ of about 200 (i.e. $\log \xi \sim 2.3$), ions would fall on a stable branch of the cooling curves. For Fe specifically, this means up to and including \ion{Fe}{xvii} ions are stable on the cooling curves corresponding to the four unobscured SEDs. But for the cooling curve of the obscured SED, only up to \ion{Fe}{viii} ions fall on the stable branch of the cooling curve. Compared to cooling curves of unobscured SEDs, the obscured version in 2013 has a remarkably extended unstable branch, where the slope is negative, and so the unstable branch covers a broader range in $\Xi$. Moreover, below a temperature of 0.1 keV, the obscured cooling curves are shifted to higher $\Xi$ and above 0.1 keV to lower $\Xi$, illustrating a more pronounced S-curve. This contrasting shape of the cooling curve is mostly induced by the relative lack of ionising EUV/soft X-ray photons in the obscured case, changing the interplay between the heating and cooling processes. Furthermore, there are some deviations between the 2013 unobscured cooling curve and those from 2000 and 2001. The main reason for this is spectral hardening of the primary power-law component in 2013 ($\Gamma = 1.6$) relative to 2000 and 2001 ($\Gamma = $1.8--1.9). This creates more emission at hard X-rays affecting the interplay between Compton heating and cooling. 
\\

Interestingly, in all five versions of the stability curves displayed in Fig. \ref{stab_curves}, the peak concentration of highly-ionised ions of \ion{Fe}{xxiii} to \ion{Fe}{xxvi} fall on an unstable branch. This means the existence of these ions is expected to be significantly diminished in photoionised gas of \ngc, regardless of obscuration effects. It is worth noting that the ionising SEDs used for the computation of the cooling curves are based on the direct accretion-powered emission from the disk as obtained from the broadband continuum modelling. So although the continuum model of the SED includes IR emission from the disk, it does not include reprocessed IR emission from the AGN dusty torus which is located further out. Therefore, enhanced cooling through inverse Comptonisation of the torus IR photons is not included. The obscurer in \ngc is located at distances of only a few light days from the central source \citep{Kaas14}, so it is expected to receive only direct disk radiation. On the other hand, the more distant warm absorbers at pc scales in \ngc \citep{Arav14} might be exposed to IR radiation from the torus, thus influencing their cooling, which is not taken into account in the calculation of the cooling curves here. The cooling curves can be used to investigate whether multiple ionisation phases of the warm absorber can co-exist in pressure equilibrium or not, since phases with overlapping values of $\Xi$ on the curves would be in pressure equilibrium. In a forthcoming paper on our campaign \citep{Ebre14} the analysis of different phases of the warm absorber and their long-term variability and position on the relevant cooling curves are investigated.

\section{Conclusions}
\label{conclusions}

In this work we have determined the intrinsic continuum of \ngc from near-infrared (8060 \AA) to hard X-ray (200 keV) energies, using stacked simultaneous \xmm, \nustar, \swift, \integral, HST COS, \wise and OCA observations from our summer 2013 campaign. We have also determined the underlying broadband continuum from \xmm observations taken in 2000 and 2001 before \ngc became obscured, which helped us in establishing the proper model for the continuum as the X-ray spectrum during the obscured epoch is more suppressed and complex. We have taken into account various NIR/optical/UV/X-ray emission and absorption processes taking place in our line of sight towards the central source, including the effects of X-ray obscuration and warm absorption, to uncover the intrinsic continuum components. We conclude that:

\begin{enumerate}

\item The soft X-ray excess in \ngc can be explained as the tail of a continuum component extending from NIR/optical/UV to soft X-ray energies, produced by Compton up-scattering of the thermal seed photons from the accretion disk, in a warm ($T_{\rm e} \approx$ 0.15~keV), optically thick ($\tau \approx 23$) corona as part of the inner disk.
\\

\item The NIR/optical/UV continuum is consistent with being composed of a single Comptonised disk component which also produces the soft X-ray excess. There is no evidence of an additional purely-thermal disk component or an additional reprocessed component from the disk.
\\

\item The nature of the underlying broadband continuum in both unobscured and obscured epochs is unchanged. However, in 2013 the primary power-law displays a harder spectrum ($\Gamma = 1.6$) than in 2000 and 2001, but this has no apparent relation to the obscuration event.
\\

\item The heavy absorption of the ionising SED EUV/soft X-ray photons by the obscurer has a significant impact on the ionisation balance of the more distant photoionised gas (warm absorbers), producing an extended thermally unstable branch on the cooling curve.
\\

\item The thermal stability analysis of photoionised gas in \ngc indicates highly-ionised ions of \ion{Fe}{xxiii} to \ion{Fe}{xxvi} fall on an unstable branch of cooling curves, regardless of the obscuration effects.

\end{enumerate}

\begin{acknowledgements}

This work is based on observations obtained with XMM-Newton, an ESA science mission with instruments and contributions directly funded by ESA Member States and the USA (NASA). This research has made use of data obtained with the NuSTAR mission, a project led by the California Institute of Technology (Caltech), managed by the Jet Propulsion Laboratory (JPL) and funded by NASA. It is also based on observations with INTEGRAL, an ESA project with instrument and science data centre funded by ESA member states (especially the PI countries: Denmark, France, Germany, Italy, Switzerland, Spain), Czech Republic, and Poland and with the participation of Russia and the USA. This work made use of data supplied by the UK Swift Science Data Centre at the University of Leicester. We thank the Chandra team for allocating the LETGS triggered observations. We thank the International Space Science Institute (ISSI) in Bern for their support and hospitality. SRON is supported financially by NWO, the Netherlands Organization for Scientific Research. This publication is supported as a project of the Nordrhein-Westf\"alische Akademie der Wissenschaften und der K\"unste in the framework of the academy program by the Federal Republic of Germany and the state Nordrhein-Westfalen. M.M. acknowledges support from NWO and the UK STFC. This work was supported by NASA through grants for HST program number 13184 from the Space Telescope Science Institute, which is operated by the Association of Universities for Research in Astronomy, Incorporated, under NASA contract NAS5-26555. M.C. acknowledges financial support from contracts ASI/INAF n.I/037/12/0 and PRIN INAF 2011 and 2012. P.-O.P. acknowledges financial support from the CNES and from the CNRS/PICS. K.C.S. acknowledges financial support from the Fondo Fortalecimiento de la Productividad Cient\'{i}fica VRIDT 2013. E.B. is supported by grants from the ISF, MoST (1163/10), and the iCORE program. S.B., G.M. and A.D.R. acknowledge INAF/PICS financial support. G.M. and F.U. acknowledge financial support from the Italian Space Agency under grant ASI/INAF I/037/12/0-011/13. B.M.P. acknowledges support from the US NSF through grant AST-1008882. G.P. acknowledges support via an EU Marie Curie Intra-European fellowship under contract no. FP-PEOPLE-2012-IEF-331095 and Bundesministerium f\"{u}r Wirtschaft und Technologie/Deutsches Zentrum f\"{u}r Luft- und Raumfahrt (BMWI/DLR, FKZ 50 OR 1408). F.U. acknowledges PhD funding from the VINCI program of the French-Italian University. M.W. acknowledges the support of a PhD studentship awarded by the UK STFC. We thank the anonymous referee for their useful suggestions and comments.

\end{acknowledgements}



\appendix
\section{Observations and data reduction}
\label{data_appendix}

\subsection{XMM-Newton}

\subsubsection*{EPIC}

The European Photon Imaging Camera (EPIC - \citealt{Stru01,Turn01}) data of \ngc were first processed using the Science Analysis System (SAS v13.0). The EPIC instruments were operating in the Small-Window mode with the thin-filter applied. Periods of high flaring background were filtered out by applying the {\tt \#XMMEA\_EP} and {\tt \#XMMEA\_EM} filterings for pn and MOS, respectively. This is done by extracting single event, high energy light curves, in order to create a set of good-time-intervals (GTIs) for use in conjunction with the internal GTI tables, to exclude intervals of flaring particle background exceeding 0.4 count/s for pn and 0.35 count/s for MOS. The EPIC spectra and lightcurves were extracted from a circular region centred on the source with a radius of $40''$. For pn the background was extracted from a nearby source-free region of radius $50''$ on the same CCD as the source, whereas for MOS a background radius of $100''$ was used from a CCD other than that of the source. For archival (2000, 2001) EPIC data, there was weak pile-up (few \% for pn); however, for the 2013 observations due to the suppressed X-ray fluxes no pile-up was detected. The single and double events were selected for both the pn ({\tt PATTERN <= 4}) and MOS ({\tt PATTERN <= 12}). The EPIC lightcurves were background-subtracted and corrected (using the {\tt epiclccorr} SAS task) for various effects on the detection efficiency. Response matrices were generated for the spectrum of each observation using the {\tt rmfgen} and {\tt arfgen} tasks. With EPIC-pn operating in the Small-Window mode, the final cleaned EPIC-pn exposure for each \xmm observation was on average about 37 ks, which is about 67\% of the duration of an observation (i.e. about 55 ks as given in Table \ref{obslog}).   
\\

The 2013--2014 EPIC-pn data of \ngc were found to be affected by larger than expected long-term degradation/evolution of the charge transfer inefficiency (CTI). This creates a shift of about 50--60~eV at 6 keV. The CTI problem had previously been found in the \mrk campaign EPIC-pn data \citep{Pont13}, and so the long-term CTI was corrected and implemented in new SAS CCFs. For the new EPIC-pn data of \ngc, further CTI corrections had to be applied and the use of the new CCFs were not sufficient to correct for all remaining CTI. We thus used an ad-hoc correcting energy shift as described in more detail in \citet{Cappi14}, in which the full data analysis and modelling of the EPIC data is reported. This energy shift results in a poor fit near the energy of the gold M-edge of the telescope mirror, therefore, the 2.0--2.4 keV region was omitted from our spectral modelling of the 2013 EPIC-pn data.

\subsubsection*{RGS}
\label{rgs_sect}

The RGS \citep{denH01} instruments were operated in the standard Spectro+Q mode for the \xmm observations of \ngc. The processing of the RGS data has been performed at a more advanced level than the standard SAS pipelines. The details of this enhanced RGS data reduction technique are reported in \citet{Kaa11b}. The procedure takes advantage of the multi-pointing mode of \xmm and utilises accurate relative calibration for the effective area of the RGS. It also incorporates an accurate absolute wavelength calibration to improve the method for stacking time-dependent RGS spectra and enhancing the efficiency of the spectral fitting. The \citet{Kaa11b} procedure was used to produce a combined RGS-1/RGS-2 spectrum and response matrix (in {\tt SPEX} format) for each \xmm observation of \ngc, including archival 2000 and 2001 observations. Also using this procedure a single stacked RGS spectrum and response matrix from the 12 \xmm observations of summer 2013 was created for spectral fitting.

\subsubsection*{OM}

The Optical Monitor (OM - \citealt{Mas01}) photometric filters were operated in the Science User Defined image/fast mode for the 2013--2014 observations and in the Image mode for the archival observations of \ngc. In each 2013--2014 observation, OM images were taken with the V, B, U, UVW1, UVM2, and UVW2 filters, with an exposure time of 4.4 ks for each image. The OM images of \ngc were processed with SAS v13.0 {\tt omichain} pipeline with the standard parameters, to apply the necessary corrections including the removal of Modulo-8 fixed pattern noise, and produce images ready for aperture photometry. The aperture photometry on each image was done using the {\tt omsource} program. For those archival datasets in which exposures in some of the six OM filters have not been taken, we used the relations derived in \citet{Meh14b} between different photometric filters from extensive \swift monitoring of \ngc, to calculate the missing fluxes in order to carry out a uniform and consistent spectral fitting of the data. Source and background regions were selected to extract count rates, with the necessary corrections applied, including those for the point spread function, coincidence losses and time-dependent sensitivity. The OM count rates were extracted from a circle of $5.0''$ centred on the source nucleus. The background was extracted from a source-free region of the same size. The OM count rates were then transformed into the standard OGIP FITS format \citep{Arna92} to be used together with their corresponding response files in X-ray spectral fitting packages. Lastly, the files were converted into the {\tt SPEX} format using the auxiliary {\tt trafo} program of the {\tt SPEX} package.
\\

In addition to the OM photometric filters, spectra with the Optical and UV OM grisms were also obtained in each 2013 \xmm observation in the Full Low mode. The exposure time of each grism spectrum is 5.0 ks. The grism data were first processed with the SAS v13.0 $\mathtt{omgchain}$ pipeline. The necessary corrections, including that for Modulo-8 fixed pattern noise and removal of scattered light features, were applied to obtain undistorted and rotated grism images. We then used the $\mathtt{omgsource}$ program to interactively identify zero and first order dispersion spectra of our source, properly define the source and background extraction regions and extract the calibrated spectrum from the grism images of each observation. For the Optical grism the extraction and calibration was extended beyond the standard results to just below $7000\ \AA$ as described in \citet{Meh11} in order to capture the ${\rm H}\alpha\, \lambda 6563$ line. Similar to the OM broadband filter data, the grism spectra were also converted into {\tt SPEX} format for simultaneous spectral modelling with other data.  

\subsection{Swift}

\subsubsection*{XRT}

For the \swift X-ray Telescope (XRT - \citealt{Burr05}) data reduction, we used the procedure detailed in \citet{Eva09}, which is an enhanced version of the standard \swift processing including several modifications. This tool is made available online on the UK Swift Science Data Centre (UKSSDC\,\footnote{\url{http://www.swift.ac.uk}}). The XRT instrument operated in the Photon Counting (PC) mode for nearly all of the \ngc observations. The XRT data have been corrected for bad pixels and effects of vignetting and PSF to produce cleaned event files using the {\tt xrtpipeline} script. The data were not significantly affected by pile-up at the observed low count rates of the source. Using the event lists, exposure maps were made, which together with the source spectrum, were used to create a corresponding Ancillary Response File (ARF) for each interval. The corrections have been applied to both spectra and lightcurves. The optimum extraction radius for data products depends on the PC-mode count rate as reported in \citet{Eva09}, which for \ngc translated most of the time into an extraction radius of 25 pixels ($59.0''$). The XRT lightcurves at different energies were constructed from each \swift snapshot observation of \ngc. The default grades of 0--12 in the PC mode were used for event selections. The corresponding response matrix files (RMF) for the \swift observations were obtained from the HEASARC Calibration Database (CALDB).

\subsubsection*{UVOT}

The \swift UV/Optical Telescope (UVOT - \citealt{Romi05}) data of \ngc from Image-mode operations were taken with the six primary photometric filters of V, B, U, UVW1, UVM2 and UVW2. The $\mathtt{uvotsource}$ tool was used to perform aperture photometry using a circular aperture radius of 5.0\arcsec. The standard instrumental corrections and calibrations according to \citet{Poo08} were applied. Any data from when \swift had bad tracking were removed. The continuous monitoring of \ngc indicated occasional exposures in UVOT filters in which the flux suddenly dropped by about 20\% or more compared to the adjacent mean; these exposures were found to be clustered at particular regions of the detector and their cause and fix is currently being investigated by the UVOT calibration team. These occasional observations have been filtered out of our analysis (about 2\% of all UVOT observations). For the purpose of spectral fitting with {\tt SPEX}, the filter count rates and the corresponding response matrices in each filter were created.
\\

In addition to the photometric exposures, there was an extensive monitoring of \ngc with the UV grism of UVOT between 1 April and 12 September 2013. The grism observations of \ngc, which were taken about every two days, were performed in the Clocked mode, which is the optimum mode for low background as well as no presence of zeroth orders in the upper left half of the detector. The Swift UVOT Grism Calibration and Software {\tt UVOTPY}\,\footnote{\url{http://www.mssl.ucl.ac.uk/~npmk/Grism}} v1 \citep{Kuin15} was used to produce the UV grism spectra of \ngc. This software is more enhanced than the standard tools available within HEASOFT. The {\tt uvotgraspcorr} program was used to aspect correct the UVOT grism images. The {\tt uvotgetspec} module of {\tt UVOTPY} was used to extract first-order spectra of \ngc and also produce the corresponding RMF response matrices. The grism spectra were then converted into {\tt SPEX} format for spectral modelling.

\subsection{NuSTAR}

The \nustar observations were reduced using the NuSTAR Data Analysis Software (NUSTARDAS) v1.3.1, which was utilised as part of HEASOFT v6.14 distribution. For instrumental calibration of our \nustar data, CALDB version v20131007 was used. The data were processed with the standard pipeline script {\tt nupipeline} to produce level 1 calibrated and level 2 cleaned event files. The data from the South Atlantic Anomaly passages have been filtered out and event files were cleaned with the standard depth correction, which significantly reduces the internal background at high energies. The source was extracted from a circular region (radius $\sim 110''$), with the background extracted from a source-free area of equal size on the same detector. Then the {\tt nuproducts} script was run to create level 3 products (spectra, lightcurves, ARF and RMF response files) for each of the two hard X-ray telescope modules (FPMA and FPMB) onboard \nustar. The spectra and corresponding response files of the two telescopes were combined for spectral modelling using the {\tt mathpha}, {\tt addrmf}, and {\tt addarf} tools of the {\tt HEASOFT} package.
 
\subsection{INTEGRAL}

The \integral IBIS and ISGRI data were reduced with the Off-line Scientific Analysis (OSA) v10.0 software. The spectra were extracted using the standard spectral extraction script {\tt ibis\_science\_analysis}. The total \integral ISGRI spectrum for each observation, and the associated RMF response and ARF ancillary files, were then generated using the OSA {\tt spe\_pick} tool. 

\subsection{Chandra LETGS}

The three \chandra observations were taken with LETGS using the HRC-S camera. The LETGS data were first reduced using the Chandra Interactive Analysis of Observations (CIAO v4.5) software to create level 1.5 event files. The rest of the processing, until the creation of final level 2 event files, was carried out using an in-house software first described by \citet{Kaa02}. This procedure follows the same steps as in the standard pipelines of CIAO, but results in improved wavelength and effective area calibration. The outcome of this procedure is the production of spectral files and their corresponding response matrices in the {\tt SPEX} format \citep{Kaa96}.

\subsection{HST COS}
\label{cos_sect}

The six HST COS observations were taken concurrently with \xmm Obs. 1, 4, 8, 11, 12 and 14. Each COS observation consists of an HST two-orbit visit. The observations through the Primary Science Aperture were taken with gratings G130M (1770~s per visit) and G160M (2215~s per visit), covering the far-UV spectral range from 1132 to 1801 \AA. In addition to the routine processing of data with the calibration pipelines of STScI, further enhanced calibrations (as described in \citealt{Kris11} for \mrk) were applied to produce the best-quality COS spectra possible. They include refined flux calibrations that take into account up-to-date adjustments to the time-dependent sensitivity of COS, improved flat-field corrections and wavelength calibration and an optimal method for combining exposures comprising a single visit. More details on the COS calibration of \ngc data are given in \citet{Kaas14}, and in \citet{Kris14}, in which the full analysis of the HST COS spectra are also reported. For the purpose of modelling the UV continuum as part of our broadband continuum modelling, we extracted the COS flux from six narrow spectral bandpasses, which are free of emission and absorption lines: 1155--1165, 1364.5--1369.5, 1470--1490, 1730--1740, 1730--1770 and 1785--1800 \AA. In order to use these measurements in our simultaneous spectral fitting with other optical/UV and X-ray data, we converted the COS fluxes into count rate units used in the OGIP spectral file format. To do this we used the time-dependent sensitivity curves as a function of wavelength, obtained for the COS data in \citet{Kris14}, which are in units of ${({\rm count}\ {\rm s}^{-1}\ {\rm pixel}^{-1}) / (\ergcm)}$, to calculate the count rate and the effective area over the six narrow bandpasses. For the G130M and G160M modes, the pixel size is about $0.010$ \AA\ and $0.012$ \AA, respectively.

\subsection{OCA}
The Observatorio Cerro Armazones (OCA) monitoring of \ngc was carried out between 20 May and 25 July 2013. For our observations the 25~cm Berlin Exoplanet Search Telescope II (BEST-II) was utilised, which is equipped with a Peltier-cooled $4096 \times 4096$ pixel Finger Lakes Imager CCD KAF-16801. It has a field-of-view of $1.7 \times 1.7$ degrees with a pixel size of 9 $\mu$m. Photometric measurements were taken in the optical Johnson B, V and R filters. For each filter, 6 dithered images were acquired, with 150~s exposure time for each individual image. There were observations on 27 days, from which 24 had exposures in B, 19 in R, and 7 in the V filter. Data reduction and processing of the OCA data was performed in the same manner as reported in \citet{Pozo13}.

\subsection{Wise Observatory}
The {\it Wise Observatory} (\wise) monitoring of NGC 5548 was done with the Centurion 46-cm telescope, and images were obtained with the standard Bessel B, V, R, and I filters. They cover the time from beginning of June to end of September 2013. There were also further monitoring from December 2013 to July 2014. The observations were done with STL-6303 CCD which gives a field of view of 70$\times$50 arcmin and each exposure time was 300~s. Standard data reduction was carried out with the Image Reduction and Analysis Facility (IRAF) packages using PSF photometry. Each night a few exposures in each filter were obtained and after confirming that there were no intra-night variations within the uncertainties, they were averaged to produce one measurement for each night. Calibration to absolute magnitudes was done in each filter using several stars in the field which their absolute magnitudes were taken from the SIMBAD database.

\end{document}